\documentclass[aps,10pt,showpacs,superscriptaddress]{revtex4}

\usepackage{times}
\usepackage{amsfonts}
\usepackage{amsmath}
\usepackage{amssymb}
\usepackage{graphicx}
\usepackage{epsfig}

\begin{document}

\title{Instabilities for a relativistic electron beam interacting with a laser irradiated plasma}
\author{Hrachya B. Nersisyan}
\email{hrachya@irphe.am}
\affiliation{Institute of Radiophysics and Electronics, 0203 Ashtarak, Armenia}
\affiliation{Centre of Strong Fields Physics, Yerevan State University, Alex Manoogian
str. 1, 0025 Yerevan, Armenia}
\author{Claude Deutsch}
\email{claude.deutsch@pgp.u-psud.fr}
\affiliation{LPGP (UMR-CNRS 8578), Universit\'{e} Paris XI, 91405 Orsay, France}
\date{\today }

\begin{abstract}
The effects of a radiation field (RF) on the unstable modes developed in relativistic electron beam--plasma
interaction are investigated assuming that $\omega_{0} >\omega_{p}$, where $\omega_{0}$ is the frequency of
the RF and $\omega_{p}$ is the plasma frequency. These unstable modes are parametrically coupled to each other
due to the RF and are a mix between two--stream and parametric instabilities. The dispersion equations are derived by the
linearization of the kinetic equations for a beam--plasma system as well as the Maxwell equations. In order to
highlight the effect of the radiation field we present a comparison of our analytical and numerical results obtained
for nonzero RF with those for vanishing RF. Assuming that the drift velocity $\mathbf{u}_{b}$ of the beam is parallel
to the wave vector $\mathbf{k}$ of the excitations two particular transversal and parallel configurations of the
polarization vector $\mathbf{E}_{0}$ of the RF with respect to $\mathbf{k}$ are considered in detail. It is shown
that in both geometries resonant and nonresonant couplings between different modes are possible. The largest growth
rates are expected at the transversal configuration when $\mathbf{E}_{0}$ is perpendicular to $\mathbf{k}$. In this
case it is demonstrated that in general the spectrum of the unstable modes in $\omega $--$k$ plane is split into two
distinct domains with long and short wavelengths, where the unstable modes are mainly sensitive to the beam or the RF
parameters, respectively. In parallel configuration, $\mathbf{E}_{0} \parallel \mathbf{k}$, and at short wavelengths
the growth rates of the unstable modes are sensitive to both beam and RF parameters remaining insensitive to the RF
at long wavelengths.
\end{abstract}

\maketitle

\section{Introduction}
\label{sec:int}

The interaction of a relativistic electron beam (REB) with a plasma is a subject of relevance for many fields of
physics ranging from inertial fusion to astrophysics \cite{ham70,pir04,gol84,aro79,die05}. This interaction is also
relevant, among others, for fast ignition scenario (FIS) \cite{tab94,deu96} where the precompressed deuterium--tritium
(DT) core of a fusion target is to be ignited by a laser--generated relativistic electron beam. The REB quickly
prompts a return current so that one eventually has to deal with a typical two--stream configuration which is subjected
to various electromagnetic instabilities. Much effort have been devoted in the last years to investigate these
instabilities \cite{ren04,bat05,tat03,fon03,sil02,hon00,bre04,bre05,bre06,bret06,bret05,bre08,cal98,hon04}, whether
it be the two--stream \cite{boh49,bun58}, the filamentation \cite{fri59} or Weibel \cite{wei59} instabilities. These
instabilities are usually analyzed by linearizing the relativistic Vlasov or fluid and Maxwell equations. Then, the
response of the linearized equation to a perturbation is investigated and one eventually finds some unstable self--excited
modes. At this stage, the orientation of the wave vector $\mathbf{k}$ of the excitations plays an important role. Choosing
a wave vector parallel to the beam velocity $\mathbf{u}_{b}$ yields the two--stream unstable modes which are of
electrostatic nature and therefore characterized by wave and electric field vectors both parallel to the beam propagation
direction. On the other hand, choosing a wave vector normal to the beam yields the purely transverse filamentation
unstable modes. They are mostly electromagnetic, purely growing, and develop preferentially in the plane normal to the
beam. The exploration of the much less investigated intermediate orientations has brought a very important result by
that showing that the strongest instability suffered by the system is eventually to be found for an oblique wave vector
\cite{bre04,bre05,bre06,bret06,bret05,bre08,cal98,blu60,fai70}. This most unstable mode is a mixture of the two--stream
and the filamentation instabilities but it is not damped as the last two ones. For example, the maximum two--stream
growth rate is reduced by a factor of $\gamma ^{-1}$ in the relativistic regime, while the most unstable mode only
decreases by a factor of $\gamma ^{-1/3}$, where $\gamma $ is the beam relativistic factor. The filamentation growth
rate varies as $\gamma ^{-1/2}$ and may be reduced, even canceled, by a transverse beam temperature \cite{sil02} whereas
the most unstable mode is quite insensitive to temperatures as long as they are nonrelativistic \cite{bre05}.

In general beam--plasma instabilities have been studied in detail for many physical situations including the interaction
of the cold, warm, inhomogeneous or anisotropic electron or ion beams with cold, warm, magnetized, unmagnetized, inhomogeneous
or anisotropic plasmas, see, e.g., Refs.~\cite{ren04,bat05,tat03,fon03,sil02,hon00,bre04,bre05,bre06,bret06,bret05,bre08,cal98,hon04}
and references therein (see also Refs.~\cite{mik74,mik92} for detailed bibliography). In this paper, we present a study
of the effects of a radiation field (RF) on the interaction of REB with a plasma. More specifically our objective is to
study the beam--plasma instabilities in a laser irradiated plasma which, to our knowledge, has not been
discussed in the literature. The principal motivations of the present paper are the researches on the topic of the FIS
for inertial confinement fusion \cite{tab94,deu96} which involves the interaction of a laser--generated REB with a hot
plasma. Although the concept of the FIS implies an overdense plasma and the propagation of a relativistic electron beam
from the border of a pre--compressed target to the dense core occurs without crossing the laser beam, the target plasma
is assumed to be parametrically excited by the RF through high harmonics generation. In addition, a promising inertial
confinement fusion scheme has been recently proposed \cite{sto96,rot01,fra10}, in which the plasma target is irradiated
simultaneously by intense laser and ion beams. In both situations the presence of the RF can dramatically change the main
features of the standard (i.e. when the laser is off) beam--plasma instabilities.

Previously the interaction of charged particles with plasma in the presence of the RF has been a subject of great
activity in the contexts of the stopping power \cite{ako97,ner99,ner11,hu11} (see also references therein) and
other processes of interest for applications in optics, solid--state, and fusion research. In particular, the
analytical calculations \cite{ako97,ner99,ner11} as well as the particle--in--cell (PIC) numerical simulations
\cite{hu11} show that the propagation and the subsequent stopping of the charged particles would be essentially
affected by the parametric excitations of the plasma target by means of laser irradiation. It is well known that
in general in the absence of the charged particle beams the laser irradiated plasma is subjected to the parametric
instabilities \cite{sil73,ale84}. Therefore in the present context it is expected that the above mentioned beam--plasma
unstable modes developed in a plasma irradiated simultaneously by laser and electron beams are parametrically coupled
to each other and are not, in principle distinguishable.

The plan of the paper is as follows. In Sec.~\ref{sec:2}, we outline the kinetic formulation for the interaction
of the REB with a laser irradiated plasma. The full electromagnetic response of the plasma is considered. The general
linear dispersion relations are derived in Sec.~\ref{sec:3}, which are studied for two particular cases in Secs.~\ref{sec:4}
and \ref{sec:5} with transversal and parallel configuration of the polarization vector of the RF with respect to
$\mathbf{k}$, respectively. Furthermore, assuming nonrelativistic laser intensities only the lowest (zero, first and
second) harmonics of the electromagnetic fields are considered. The obtained dispersion equations are investigated
numerically in Sec.~\ref{sec:6}. The results are summarized in Sec.~\ref{sec:7}, which also includes discussion and
outlook. In Appendix~\ref{sec:app0} we consider the standard (in the absence of the RF) stable and unstable modes of
the beam--plasma system in a cold--fluid approximation. The asymptotic behavior of the frequencies and the growth rates
of these modes at large and small $k$ are considered. In Appendix~\ref{sec:app1}, we provide some technical details for
an evaluation of the sum containing Bessel functions. An equation describing the evolution of the amplitude of the
parametrically excited plasma waves and involving all excited harmonics is derived and discussed in Appendix~\ref{sec:app2}.

\section{Theoretical background}
\label{sec:2}

In this section, we consider the main aspects of the interaction of the relativistic electron beam (REB) with a
homogeneous, collisionless, and unmagnetized plasma in the presence of high--frequency radiation field (RF),
$\mathbf{E}_{0}(t)=\mathbf{E}_{0}\sin (\omega _{0}t)$. Here $\mathbf{E}_{0}$ and $\omega _{0}$ are the amplitude
and the frequency of the RF. The problem is formulated using the perturbative treatment, and includes the effects
of the RF in a self--consistent way. The RF is treated in the long--wavelength limit, and the plasma particles
(electrons and ions) are considered nonrelativistic. These are good approximations provided that (1) the wavelength
of the RF ($\lambda _{0}=2\pi c/\omega _{0}$) is much larger than the Debye screening length $\lambda _{\mathrm{D}}%
= v_{\mathrm{th}}/\omega_{p}$ with $v_{\mathrm{th}}$ the thermal velocity of the electrons and $\omega _{p}$ the
plasma frequency, and (2) the \textquotedblright quiver velocity\textquotedblright of the electrons in the RF
($v_{E}=eE_{0}/m\omega _{0}$) is much smaller than the speed of light $c$. These conditions can be alternatively
written as (1) $\omega_{0}/\omega_{p} \ll 2\pi c/v_{\mathrm{th}}$, (2) $I_{L} \ll\frac{1}{2} n_{e} mc^{3}%
(\omega_{0}/\omega_{p})^{2}$, where $I_{L}=cE_{0}^{2}/8\pi$ is the RF intensity. As an estimate in the case of dense
plasma, with electron density $n_{e} =10^{23}$ cm$^{-3}$, we get $\frac{1}{2} n_{e} mc^{3} \simeq 1.2\times%
10^{20}$ W/cm$^{2}$. Thus the limits (1) and (2) are well above the values obtained with currently available
high--power RF sources, and so the approximations are well justified. Furthermore we consider an underdense
plasma with $\omega_{0} >\omega_{p}$.

The dynamics of the beam--plasma system is governed by the relativistic and
nonrelativistic Vlasov kinetic equations for the distribution functions of
the REB $f_{b}(\mathbf{r},\mathbf{v},t)$ and the plasma $f_{\alpha }(\mathbf{%
r},\mathbf{v},t)$ (where $\alpha =e,i$ indicates the plasma species),
respectively, as well as by the Maxwell equations for the electromagnetic
fields. Thus,
\begin{equation}
\frac{\partial f_{\alpha ,b}}{\partial t}+\mathbf{v}\cdot \frac{\partial
f_{\alpha ,b}}{\partial \mathbf{r}}+e_{\alpha ,b}\left\{ \mathbf{E}%
_{0}\left( t\right) +\mathbf{E}+\frac{1}{c}\left[ \mathbf{v}\times \mathbf{B}%
\right] \right\} \cdot \frac{\partial f_{\alpha ,b}}{\partial \mathbf{p}}=0,
\label{eq:1}
\end{equation}%
where $e_{\alpha }$ and $e_{b}$ are the charges of the plasma and beam
particles, respectively, $\mathbf{E}$\ and $\mathbf{B}$\ are the
self--consistent electromagnetic fields which are determined by the Maxwell
equations,
\begin{eqnarray}
&&\nabla \times \mathbf{E} =-\frac{1}{c}\frac{\partial \mathbf{B}}{\partial t%
} , \qquad \nabla \times \mathbf{B} = \frac{4\pi }{c}\left( \sum_{\alpha }%
\mathbf{j}_{\alpha }+\mathbf{j}_{b}\right) +\frac{1}{c}\frac{\partial
\mathbf{E}}{\partial t} ,  \label{eq:2} \\
&&\nabla \cdot \mathbf{E} =4\pi \left( \sum_{\alpha }e_{\alpha }n_{\alpha
}+e_{b}n_{b}\right) , \qquad \nabla \cdot \mathbf{B}=0 .  \label{eq:3}
\end{eqnarray}%
Here $n_{\alpha }(\mathbf{r},t)$ and $n_{b}(\mathbf{r},t)$ are the densities for the plasma species $\alpha $ and for the
relativistic beam, respectively, $\mathbf{j}_{\alpha }(\mathbf{r},t)$ and $%
\mathbf{j}_{b}(\mathbf{r},t)$ are the corresponding currents induced in
plasma and beam, respectively,
\begin{equation}
n_{\alpha ,b}\left( \mathbf{r},t\right) =\int f_{\alpha ,b}\left( \mathbf{r},%
\mathbf{p},t\right) d\mathbf{p},\qquad \mathbf{j}_{\alpha ,b}\left( \mathbf{r%
},t\right) =e_{\alpha ,b}\int \mathbf{v}f_{\alpha ,b}\left( \mathbf{r},%
\mathbf{p},t\right) d\mathbf{p} .
\label{eq:4}
\end{equation}

As mentioned above we consider a nonrelativistic plasma and in Eq.~%
\eqref{eq:1} for the distribution function $f_{\alpha }$ the momentum
connects linearly to the particle velocity, $\mathbf{p}=m_{\alpha }\mathbf{v}
$, where $m_{\alpha }$ is the mass of the plasma species $\alpha $. Equation~%
\eqref{eq:1} for the distribution function $f_{b}$ is relativistic and $%
\mathbf{p}=m_{b}\gamma \mathbf{v}$ in this case, where $\gamma
=(1-v^{2}/c^{2})^{-1/2}$ and $m_{b}$ are the relativistic factor and the
rest mass of the beam particles. Moreover, we consider an ultrarelativistic
electron beam with $\gamma _{b}=(1-u_{b}^{2}/c^{2})^{-1/2}\gg 1$, where $%
u_{b}$ is the beam drift velocity, and therefore the influence of the RF $%
\mathbf{E}_{0}(t)$ on the beam is ignored in the kinetic equation %
\eqref{eq:1} for the distribution function $f_{b}(\mathbf{r},\mathbf{p},t)$.

For sufficiently small perturbations, we assume $f_{\alpha ,b}=f_{0\alpha
,b}+f_{1\alpha ,b}$ (with $f_{1\alpha ,b}\ll f_{0\alpha ,b}$), where $%
f_{0\alpha }$ and $f_{0b}$\ are the equilibrium distribution functions of
the plasma species $\alpha $ and the beam in an unperturbed state,
respectively. The solution of the linearized kinetic equation \eqref{eq:1}
for the relativistic beam, when the RF $\mathbf{E}_{0}(t)$ in Eq.~%
\eqref{eq:1} is neglected, is well known. This standard calculation is
explained at length in a number of plasma physics textbooks, \cite%
{ale84,ich73,cle90,dav83}, and we just here mention the final result. In
terms of the Fourier transformed quantities the solution of the kinetic
equation reads
\begin{equation}
f_{1b}\left( \mathbf{k},\omega ,\mathbf{p}\right) =ie_{b}E_{j}\left( \mathbf{%
k},\omega \right) \left[ \delta _{ij}\left( 1-\frac{k_{s}v_{s}}{\omega }%
\right) +\frac{k_{i}v_{j}}{\omega }\right] \frac{\partial f_{0b}\left(
\mathbf{p}\right) }{\partial p_{i}}\frac{1}{\mathbf{k}\cdot \mathbf{v}%
-\omega -i0} .
\label{eq:5}
\end{equation}%
Here $\mathbf{E}(\mathbf{k},\omega )$\ and $f_{1b}(\mathbf{k},\omega ,%
\mathbf{p})$\ are the Fourier transforms of the electric field and the beam
distribution function, respectively, with respect to variables $\mathbf{r}$
and $t$, $\delta _{ij}$ is the unit tensor of rank three. Note that the
positive infinitesimal $+i0$ in Eq.~\eqref{eq:5} guarantees the causality of
the response.

The perturbations of the densities and the currents induced in the plasma
and the beam are determined from Eq.~\eqref{eq:4}. The Fourier transforms of
these quantities are then given by
\begin{equation}
n_{1\alpha ,b}\left( \mathbf{k},\omega \right) =\int f_{1\alpha ,b}\left(
\mathbf{k},\omega ,\mathbf{p}\right) d\mathbf{p},\qquad \mathbf{j}_{1\alpha
,b}\left( \mathbf{k},\omega \right) =e_{\alpha ,b}\int \mathbf{v}f_{1\alpha
,b}\left( \mathbf{k},\omega ,\mathbf{p}\right) d\mathbf{p} .
\label{eq:6}
\end{equation}%
Substituting the distribution function~\eqref{eq:5} into these quantities
and integrating by parts with the help of the relation $\partial
v_{i}/\partial p_{j}=(1/m_{b}\gamma )(\delta _{ij}-v_{i}v_{j}/c^{2})$ (where
$\gamma ^{2}(p)=1+p^{2}/m_{b}^{2}c^{2}$) for the induced current and the
density of the beam we obtain
\begin{eqnarray}
&&j_{1b,i}\left( \mathbf{k},\omega \right) =\sigma _{b,ij}\left( \mathbf{k}%
,\omega \right) E_{j}\left( \mathbf{k},\omega \right) ,  \label{eq:7} \\
&&e_{b}n_{1b}\left( \mathbf{k},\omega \right) =\frac{ie_{b}^{2}}{m_{b}\omega
}\mathbf{E}\left( \mathbf{k},\omega \right) \cdot \int \left[ \mathbf{k}+%
\mathbf{v}\frac{k^{2}-\left( \mathbf{k\cdot v}\right) \omega /c^{2}}{\omega -%
\mathbf{k}\cdot \mathbf{v}+i0}\right] \frac{f_{0b}\left( \mathbf{p}\right) d%
\mathbf{p}}{\gamma \left( p\right) \left( \omega -\mathbf{k}\cdot \mathbf{v}%
+i0\right) },  \label{eq:8}
\end{eqnarray}%
where $\sigma _{b,ij}(\mathbf{k},\omega )$ is the conductivity tensor of the
relativistic charged particle beam,
\begin{equation}
\sigma _{b,ij}\left( \mathbf{k},\omega \right) =\frac{ie_{b}^{2}}{%
m_{b}\omega }\int \left[ \delta _{ij}+\frac{v_{i}k_{j}+k_{i}v_{j}}{\omega -%
\mathbf{k}\cdot \mathbf{v}+i0}+v_{i}v_{j}\frac{k^{2}-\omega ^{2}/c^{2}}{%
\left( \omega -\mathbf{k}\cdot \mathbf{v}+i0\right) ^{2}}\right] \frac{%
f_{0b}\left( \mathbf{p}\right) d\mathbf{p}}{\gamma \left( p\right) }.
\label{eq:9}
\end{equation}

Consider now the solution of the kinetic equation~\eqref{eq:1} for the
plasma electrons and ions in the presence of the high--frequency RF. In an
unperturbed state (i.e. neglecting the self--consistent electromagnetic
fields $\mathbf{E}$ and $\mathbf{B}$ in Eq.~\eqref{eq:1} and assuming the
homogeneous initial state) the distribution function satisfies the equation
\begin{equation}
\frac{\partial f_{0\alpha }}{\partial t}+e_{\alpha }\mathbf{E}_{0}\sin
\left( \omega _{0}t\right) \cdot \frac{\partial f_{0\alpha }}{\partial
\mathbf{p}}=0
\label{eq:10}
\end{equation}%
which yields the equilibrium distribution function for the plasma species $\alpha $,
\begin{equation}
f_{0\alpha }\left( \mathbf{p},t\right) =F_{\alpha }\left( \mathbf{p}%
+m_{\alpha }\mathbf{v}_{E\alpha }\cos \left( \omega _{0}t\right) \right) .
\label{eq:11}
\end{equation}%
Here $F_{\alpha }(\mathbf{p})$ is an arbitrary function. Below we will
assume that this function is isotropic in momentum space. $\mathbf{v}%
_{E\alpha }=e_{\alpha }\mathbf{E}_{0}/m_{\alpha }\omega _{0}$, $\mathbf{a}%
_{\alpha }=e_{\alpha }\mathbf{E}_{0}/m_{\alpha }\omega _{0}^{2}$ are quiver
velocity and the oscillation amplitude of the plasma species $\alpha $,
respectively, driven by the RF.

Next we consider the linearized kinetic equation for the plasma species $%
\alpha $. Introducing the Fourier transforms $f_{1\alpha }(\mathbf{k},%
\mathbf{p},t)$, $\mathbf{E}(\mathbf{k},t)$ and $\mathbf{B}(\mathbf{k},t)$\
with respect to the coordinate $\mathbf{r}$ the linearized kinetic equation
reads
\begin{eqnarray}
&&\left[ \frac{\partial }{\partial t}+i\left( \mathbf{k}\cdot \mathbf{v}%
\right) +e_{\alpha }\mathbf{E}_{0}\sin \left( \omega _{0}t\right) \cdot
\frac{\partial }{\partial \mathbf{p}}\right] f_{1\alpha }\left( \mathbf{k},%
\mathbf{p},t\right)  \label{eq:12} \\
&& =-e_{\alpha }\left[ \mathbf{E}\left( \mathbf{k},t\right) +\frac{1}{c}%
\left[ \mathbf{v}\times \mathbf{B}\left( \mathbf{k},t\right) \right] \right]
\cdot \frac{\partial f_{0\alpha }}{\partial \mathbf{p}} .  \nonumber
\end{eqnarray}%
In order to solve Eq.~\eqref{eq:12} it is convenient to introduce a new
unknown function $\Psi _{\alpha }$\ via relation
\begin{equation}
f_{1\alpha }\left( \mathbf{k},\mathbf{p},t\right) =e^{i\zeta _{\alpha }\sin
\left( \omega _{0}t\right) }\Psi _{\alpha }\left( \mathbf{k},\mathbf{P}%
,t\right) ,
\label{eq:13}
\end{equation}%
where $\zeta _{\alpha }=(\mathbf{k}\cdot \mathbf{a}_{\alpha })$ and $\mathbf{%
P}=\mathbf{p}+m_{\alpha }\mathbf{v}_{E\alpha }\cos (\omega _{0}t)$.
Substituting this relation into Eq.~\eqref{eq:12} it is easy to see that the
obtained equation for the unknown function $\Psi _{\alpha }$\ constitutes an
equation with periodic coefficients where the role of the dynamic momentum
is now played by the quantity $\mathbf{P}$. Therefore we introduce the
decomposition
\begin{equation}
Q(\mathbf{k},\mathbf{p},t) =\int_{-\infty }^{\infty }d\omega e^{-i\omega t}\sum_{n=-\infty }^{\infty }
Q^{(n)}(\mathbf{k},\omega ,\mathbf{p}) e^{-in\omega _{0}t} ,
\label{eq:14}
\end{equation}%
where $Q(\mathbf{k},\mathbf{p},t)$ represents one of the
quantities $\Psi _{\alpha }(\mathbf{k},\mathbf{p},t)$, $\mathbf{E}(\mathbf{k}%
,t)$ and $\mathbf{B}(\mathbf{k},t)$ and $Q^{(n)}(\mathbf{k},\omega ,\mathbf{p%
})$ is the corresponding amplitude of the $n$th harmonic. From Maxwell
equation we express the magnetic field through the electric field. In terms
of the amplitudes of the $n$th harmonics (see definition given by Eq.~%
\eqref{eq:14}) this relation is given by
\begin{equation}
\mathbf{B}^{(n)}(\mathbf{k},\omega ) =\frac{c}{\omega +n\omega _{0}}[\mathbf{k}\times
\mathbf{E}^{(n)}(\mathbf{k},\omega )].
\label{eq:15}
\end{equation}%
Also in the kinetic equation derived for $\Psi _{\alpha }^{(n)}$ taking into
account the isotropy of the equilibrium distribution function $F_{\alpha }$
of the plasma particles we neglect the term
\begin{equation}
\left[ k_{i}v_{j}-\delta _{ij}\left( \mathbf{k}\cdot \mathbf{v}\right) %
\right] \cdot \frac{\partial F_{\alpha }}{\partial p_{i}}=0
\label{eq:16}
\end{equation}%
and using the Fourier series representation of the exponential function \cite%
{gra80} (see also Eq.~\eqref{eq:bp1}) for the amplitude of the $n$th
harmonic of the distribution function we finally obtain
\begin{eqnarray}
&&\Psi _{\alpha }^{(n)}\left( \mathbf{k},\omega ,\mathbf{p}\right) =-\frac{%
ie_{\alpha }}{\omega +n\omega _{0}-\mathbf{k}\cdot \mathbf{v}+i0}\frac{%
\partial F_{\alpha }\left( \mathbf{p}\right) }{\partial p_{i}}  \label{eq:17}
\\
&&\times \sum_{\ell =-\infty }^{\infty }J_{n-\ell }\left( \zeta _{\alpha
}\right) \left[ \delta _{ij}+\frac{\left( n-\ell \right) \omega _{0}}{\omega
+\ell \omega _{0}}\left( \delta _{ij}-\frac{k_{i}a_{\alpha ,j}}{\zeta
_{\alpha }}\right) \right] E_{j}^{(\ell )}\left( \mathbf{k},\omega \right) . \nonumber
\end{eqnarray}%
Here $J_{n}$ is the Bessel function of the $n$th order. Throughout this paper the upper indices given in the
parentheses indicate the harmonic number while the lower indices determine the components of the vectors and
tensors.

The amplitudes of the harmonics of the induced current and the charge
density in a plasma are obtained from Eqs.~\eqref{eq:6}, \eqref{eq:13}, %
\eqref{eq:14} and \eqref{eq:17}. Straightforward calculations yield
\begin{eqnarray}
&&j_{r}^{(s)}\left( \mathbf{k},\omega \right) =\sum_{\ell =-\infty }^{\infty
}\sigma _{rj}^{(s\ell )}\left( \mathbf{k},\omega \right) E_{j}^{(\ell
)}\left( \mathbf{k},\omega \right) ,  \label{eq:18} \\
&&e_{\alpha }n_{\alpha }^{(s)}\left( \mathbf{k},\omega \right) =-\frac{ik}{%
4\pi }\sum_{n=-\infty }^{\infty }J_{n-s}\left( \zeta _{\alpha }\right)
\delta \varepsilon _{\alpha \parallel }\left( n\right)  \label{eq:19} \\
&&\times \sum_{\ell =-\infty }^{\infty }\frac{\omega +n\omega _{0}}{\omega
+\ell \omega _{0}}J_{n-\ell }\left( \zeta _{\alpha }\right) [\boldsymbol{%
\chi }^{(\ell n)}\left( \mathbf{k},\omega \right) \cdot \mathbf{E}^{(\ell
)}\left( \mathbf{k},\omega \right) ]  \nonumber
\end{eqnarray}%
for the Fourier transforms of the $s$th harmonics of the current and the
charge density, respectively, with
\begin{equation}
\boldsymbol{\chi }^{(\ell n)}\left( \mathbf{k},\omega \right) =\frac{\mathbf{%
k}}{k}+\frac{\left( \ell -n\right) \omega _{0}}{\omega +n\omega _{0}}\frac{k%
}{\zeta _{\alpha }}\mathbf{a}_{\alpha } .
\label{eq:20}
\end{equation}%
Here we have introduced the conductivity tensor $\sigma _{rj}^{(s\ell )}(\mathbf{k},\omega )$
\begin{eqnarray}
&&\sigma _{rj}^{(s\ell )}\left( \mathbf{k},\omega \right) =\frac{1}{4\pi i}%
\sum_{\alpha }\sum_{n=-\infty }^{\infty }\frac{\left( \omega +n\omega
_{0}\right) ^{2}}{\omega +\ell \omega _{0}}J_{n-\ell }\left( \zeta _{\alpha
}\right) J_{n-s}\left( \zeta _{\alpha }\right)  \label{eq:21} \\
&&\times \left[ \left( \delta _{jr}-\frac{k_{j}k_{r}}{k^{2}}\right) \delta
\varepsilon _{\alpha \perp }\left( n\right) +\chi _{j}^{(\ell n)}\left(
\mathbf{k},\omega \right) \chi _{r}^{(sn)}\left( \mathbf{k},\omega \right)
\delta \varepsilon _{\alpha \parallel }\left( n\right) \right]  \nonumber
\end{eqnarray}%
and the abbreviations $\delta \varepsilon _{\alpha \parallel ;\perp
}(n)\equiv \delta \varepsilon _{\alpha \parallel ;\perp }(k,\omega +n\omega
_{0})$, where $\delta \varepsilon _{\alpha \parallel }(k,\omega )$\ and $%
\delta \varepsilon _{\alpha \perp }(k,\omega )$\ are the partial
contributions of the plasma species $\alpha $ to the longitudinal and
transversal (with respect to the wave vector $\mathbf{k}$) dielectric
functions (see, e.g., \cite{ich73}), respectively,
\begin{eqnarray}
&&\delta \varepsilon _{\alpha \parallel }\left( k,\omega \right) =\frac{4\pi
e_{\alpha }^{2}}{k^{2}}k_{i}\int \frac{\partial F_{\alpha }\left( \mathbf{p}%
\right) }{\partial p_{i}}\frac{d\mathbf{p}}{\omega -\mathbf{k}\cdot \mathbf{v%
}+i0},  \label{eq:22} \\
&&\delta \varepsilon _{\alpha \perp }\left( k,\omega \right) =\frac{2\pi
e_{\alpha }^{2}}{\omega }\int \left[ v_{i}-\frac{\left( \mathbf{k}\cdot
\mathbf{v}\right) k_{i}}{k^{2}}\right] \frac{\partial F_{\alpha }\left(
\mathbf{p}\right) }{\partial p_{i}}\frac{d\mathbf{p}}{\omega -\mathbf{k}%
\cdot \mathbf{v}+i0}.  \label{eq:23}
\end{eqnarray}%
Note that since the equilibrium distribution function $F_{\alpha }$ is
isotropic the partial dielectric functions $\delta \varepsilon _{\alpha
\parallel ;\perp }(k,\omega )$ are also isotropic, i.e. they do not depend
on the direction of the wave vector $\mathbf{k}$. The obtained expressions %
\eqref{eq:18}--\eqref{eq:23} with Eqs.~\eqref{eq:7}--\eqref{eq:9} as well as
the Maxwell equations \eqref{eq:2} and \eqref{eq:3} written in the Fourier
space completely determine the electromagnetic response in the beam--plasma
system in the presence of the RF. Using this system of equations the general
dispersion equations are derived in the next section.

We would like to close this section with the following two remarks. First,
the distribution function $f_{1b}(\mathbf{k},\omega ,\mathbf{p})$ of the
beam particles given by Eq.~\eqref{eq:5} as well as the induced current %
\eqref{eq:7} and the charge density \eqref{eq:8} are determined by the
Fourier transform $\mathbf{E}(\mathbf{k},\omega )$ of the electric field. In
contrast to this case the distribution function (Eq.~\eqref{eq:13} with Eqs.~%
\eqref{eq:14} and \eqref{eq:17}), the induced current (Eq.~\eqref{eq:18})
and the density (Eq.~\eqref{eq:19}) of the plasma are determined by the
Fourier transform of the amplitude of the harmonics of the electric field, $%
\mathbf{E}^{(n)}(\mathbf{k},\omega )$. From Eq.~\eqref{eq:14} it is
straightforward to deduce the connection between the Fourier transforms $%
\mathbf{E}(\mathbf{k},\omega )$ and $\mathbf{E}^{(n)}(\mathbf{k},\omega )$.
Changing the integration variable in each term of summation in Eq.~%
\eqref{eq:14} according to $\omega +n\omega _{0}\rightarrow \omega $ we
obtain
\begin{equation}
\mathbf{E}\left( \mathbf{k},\omega \right) =\sum_{n=-\infty }^{\infty }%
\mathbf{E}^{(n)}\left( \mathbf{k},\omega -n\omega _{0}\right) .
\label{eq:24}
\end{equation}%
Thus, $\mathbf{E}(\mathbf{k},\omega )$ is the sum of all harmonics $\mathbf{E%
}^{(n)}(\mathbf{k},\omega )$ with shifted frequencies $\omega \pm n\omega
_{0}$. Second, assuming an ultrarelativistic beam for derivation of the
distribution function \eqref{eq:5} we have neglected the RF in the kinetic
equation \eqref{eq:1} for $f_{b}$. And as a consequence the perturbation of
the beam distribution function is determined by $\mathbf{E}(\mathbf{k}%
,\omega )$. Although the RF is not directly involved in the kinetic equation %
\eqref{eq:1} for $f_{b}$, it affects this distribution function via the
self--consistent electric field $\mathbf{E}(\mathbf{k},\omega )$ containing
all harmonics produced by the RF (see Eq.~\eqref{eq:24}).

\section{Dispersion equation}
\label{sec:3}

In this section using the expressions derived for the induced currents in
the beam and plasma we consider the dispersion equation of the waves excited
in a plasma by the relativistic beam of charged particles. For this purpose
we employ the Maxwell equations~\eqref{eq:2}. Introducing Fourier transforms
of the electric field and the currents according to Eq.~\eqref{eq:14} and
excluding the magnetic field from these equations by means of Eq.~\eqref{eq:15} from Eqs.~\eqref{eq:2},
\eqref{eq:7}--\eqref{eq:9} and \eqref{eq:18}--\eqref{eq:21} for the components of the amplitude of the $n$th
harmonic of the electric field we obtain
\begin{eqnarray}
&&\left\{ \delta _{rj}\left[ k^{2}-\frac{\left( \omega +n\omega _{0}\right)
^{2}}{c^{2}}\right] -k_{r}k_{j}\right\} E_{j}^{(n)}\left( \mathbf{k},\omega
\right) =\frac{4\pi i\left( \omega +n\omega _{0}\right) }{c^{2}}
\label{eq:25} \\
&&\times \left[ \sum_{\ell =-\infty }^{\infty }\sigma _{rj}^{(n\ell )}\left(
\mathbf{k},\omega \right) E_{j}^{(\ell )}\left( \mathbf{k},\omega \right)
+\delta _{n0}\sigma _{b,rj}\left( \mathbf{k},\omega \right) E_{j}\left(
\mathbf{k},\omega \right) \right] .  \nonumber
\end{eqnarray}%
Here the conductivity tensors of the plasma $\sigma _{rj}^{(n\ell )}(\mathbf{%
k},\omega )$ and the beam $\sigma _{b,rj}(\mathbf{k},\omega )$ are
determined by Eqs.~\eqref{eq:21} and \eqref{eq:9}, respectively. It is seen
that in the right--hand side of Eq.~\eqref{eq:25} the beam current vanishes
for any nonzero harmonic number, $n\neq 0$.

Before starting the systematic investigation of the general dispersion
equation for the beam--plasma system in the presence of the RF it is
constructive consider briefly two limiting cases of Eq.~\eqref{eq:25}.
First, at the vanishing RF (i.e. at $\zeta _{\alpha }\rightarrow 0$) from
Eq.~\eqref{eq:21} it is straightforward to calculate the conductivity tensor
$\sigma _{ij}^{(n\ell )}(\mathbf{k},\omega )$ of the plasma which reads in
this limit
\begin{equation}
\sigma _{ij}^{(n\ell )}\left( \mathbf{k},\omega \right) =\delta _{n\ell }%
\frac{\omega +n\omega _{0}}{4\pi i}\left[ \varepsilon _{ij}\left( n\right)
-\delta _{ij}\right] \equiv \delta _{n\ell }\sigma _{ij}\left( n\right) .
\label{eq:26}
\end{equation}%
Here we have introduced (as above) the abbreviations $\sigma _{ij}(n)\equiv
\sigma _{ij}(\mathbf{k},\omega +n\omega _{0})$, $\varepsilon _{ij}(n)\equiv
\varepsilon _{ij}(\mathbf{k},\omega +n\omega _{0})$, $\varepsilon
_{\parallel ;\perp }(n)\equiv \varepsilon _{\parallel ;\perp }(k,\omega
+n\omega _{0})$, and $\sigma _{ij}(\mathbf{k},\omega )$\ and
\begin{equation}
\varepsilon _{ij}\left( \mathbf{k},\omega \right) =\frac{k_{i}k_{j}}{k^{2}}%
\varepsilon _{\parallel }\left( k,\omega \right) +\left( \delta _{ij}-\frac{%
k_{i}k_{j}}{k^{2}}\right) \varepsilon _{\perp }\left( k,\omega \right)
\label{eq:27}
\end{equation}%
are the conductivity and the dielectric tensors of an isotropic plasma,
respectively, with longitudinal ($\varepsilon _{\parallel }(k,\omega )$) and
transversal ($\varepsilon _{\perp }(k,\omega )$) dielectric functions (see,
e.g., Refs.~\cite{ale84,ich73}),
\begin{equation}
\varepsilon _{\parallel ,\perp }\left( k,\omega \right) =1+\sum_{\alpha
}\delta \varepsilon _{\alpha \parallel ,\perp }\left( k,\omega \right) .
\label{eq:28}
\end{equation}%
Substituting Eq.~\eqref{eq:26} into Eq.~\eqref{eq:25} we obtain that $%
E_{j}^{(n)}(\mathbf{k},\omega )=\delta _{n0}E_{j}(\mathbf{k},\omega )$ and
Eq.~\eqref{eq:24} is fulfilled automatically. Thus, after some
simplifications we arrive at
\begin{equation}
\left\{ k^{2}\delta _{ij}-k_{i}k_{j}-\frac{\omega ^{2}}{c^{2}}\left[
\varepsilon _{ij}\left( \mathbf{k},\omega \right) +\delta \varepsilon
_{b,ij}\left( \mathbf{k},\omega \right) \right] \right\} E_{j}\left( \mathbf{%
k},\omega \right) =0 .
\label{eq:29}
\end{equation}%
Here $\varepsilon _{ij}(\mathbf{k},\omega )+\delta \varepsilon _{b,ij}(\mathbf{k},\omega )$ is the total dielectric
tensor of the beam--plasma system, where $\delta \varepsilon _{b,ij}(\mathbf{k},\omega )=(4\pi i/\omega)\sigma _{b,ij}%
(\mathbf{k},\omega )$ is the partial contribution of the beam to the total dielectric tensor of the system. Equating the
determinant of the system of linear equations \eqref{eq:29} to zero yields the general dispersion equation for the
beam--plasma system. In the last years this equation has been studied in detail for arbitrary orientation of the electron
beam propagation direction with respect to the wave vector $\mathbf{k}$, see, e.g.,
Refs.~\cite{tat03,fon03,sil02,hon00,bre04,bre05,bre06,bret06,bret05,bre08,cal98,hon04}. In particular, assuming cold,
homogeneous and monochromatic charged particle beam with unperturbed distribution function $f_{0b}(\mathbf{p})=n_{b}\delta%
(\mathbf{p}-\mathbf{p}_{b})$, where $n_{b}$ is the beam density, from Eq.~\eqref{eq:9} one obtains
\begin{equation}
\delta \varepsilon _{b,ij}\left( \mathbf{k},\omega \right) =-\frac{\omega
_{b}^{2}}{\gamma _{b}\omega ^{2}}\left[ \delta _{ij}+\frac{%
u_{bi}k_{j}+u_{bj}k_{i}}{\omega -\mathbf{k}\cdot \mathbf{u}_{b}}+\frac{%
u_{bi}u_{bj}\left( k^{2}-\omega ^{2}/c^{2}\right) }{\left( \omega -\mathbf{k}%
\cdot \mathbf{u}_{b}\right) ^{2}}\right] .
\label{eq:30}
\end{equation}%
Here $\mathbf{p}_{b}=m_{b}\gamma _{b}\mathbf{u}_{b}$, $\gamma
_{b}=(1-u_{b}^{2}/c^{2})^{-1/2}$ and $\omega _{b}^{2}=4\pi
n_{b}e_{b}^{2}/m_{b}$ are the relativistic factor and the plasma frequency
of the beam, respectively. At this stage it is convenient to represent the
vectors (particularly the electric field) in the form of an expansion in the
components parallel, $A_{\parallel }=(\mathbf{k}\cdot \mathbf{A})/k$, and
perpendicular, $\mathbf{A}_{\perp }=\mathbf{A}-(\mathbf{k}/k)A_{\parallel }$%
, to the wave vector $\mathbf{k}$. In particular, choosing a wave vector $%
\mathbf{k}$ parallel to the beam velocity $\mathbf{u}_{b}$ yields the
two--stream (TS) unstable modes which are of electrostatic nature with $%
E_{\parallel }(\mathbf{k},\omega )\neq 0$ and $\mathbf{E}_{\perp }(\mathbf{k}%
,\omega )=0$. Introducing the longitudinal dielectric function of the beam
by means of the relation,
\begin{equation}
\delta \varepsilon _{b\parallel }\left( \mathbf{k},\omega \right) =\frac{%
k_{i}k_{j}}{k^{2}}\delta \varepsilon _{b,ij}\left( \mathbf{k},\omega \right)
=-\frac{\omega _{b}^{2}}{\gamma _{b}^{3}\left( \omega -ku_{b\parallel
}\right) ^{2}} ,
\label{eq:31}
\end{equation}%
the dispersion equation in this case for a beam--plasma system then reads
\begin{equation}
\mathcal{D}_{\parallel }\left( k,\omega \right) \equiv \varepsilon
_{\parallel }\left( k,\omega \right) +\delta \varepsilon _{b\parallel
}\left( k,\omega \right) =0 .
\label{eq:32}
\end{equation}%
On the other hand, choosing $\mathbf{k}$ normal to the beam velocity $\mathbf{u}_{b}$\ yields the purely
transverse (electromagnetic) filamentation unstable modes. It should be emphasized that we have considered
above an infinite beam of charged particles and as a consequence the plasma return current is not involved
in Eqs.~\eqref{eq:30}--\eqref{eq:32} in self--consistent manner. This is not, however, a strong limitation
of the present treatment. For instance, the drift velocity $\mathbf{u}_{e}$ of the plasma return flow can
be deduced from the beam current neutralization condition, $n_{e}\mathbf{u}_{e} \simeq -n_{b}\mathbf{u}_{b}$.
Then within a cold--fluid model the return current is included by adding in Eqs.~\eqref{eq:30} and \eqref{eq:31}
the similar terms but with a flow velocity $\mathbf{u}_{e}$ and plasma density $n_{e}$ \cite{bre04,bre05,bre06}.

Second, in the case of the absence of the external beam the last term in the right--hand side of Eq.~\eqref{eq:25}
vanishes. Then the remaining infinite system of equations for the electric field harmonics represents the electromagnetic
response of the plasma to the RF. In general the longitudinal and transversal components of the electric field are
coupled parametrically and the excitations are a mixture of both types of modes. Previously the parametrically unstable
modes have been studied in detail both for the electrostatic \cite{ali66} (see also Refs.~\cite{sil73,ale84}) and
electromagnetic \cite{gor66,ali79} excitations. The purely electrostatic excitations with $\mathbf{E}_{\perp }^{(\ell )}=0$
are possible when the polarization vector of the laser radiation is parallel to the wave vector $\mathbf{k}$
($\mathbf{E}_{0\perp }=0$). In this case the plasma electrons and ions are driven by the laser field only in the
direction of $\mathbf{k}$.

To illustrate the problem of the charged particles beam--plasma instabilities developed in a laser irradiated plasma,
we consider below two examples when the polarization vector $\mathbf{E}_{0}$ of the RF is perpendicular (Sec.~\ref{sec:4})
or parallel (Sec.~\ref{sec:5}) to the wave vector $\mathbf{k}$. We consider an infinite and cold beam of charged particles
of velocity $\mathbf{u}_{b}$ aligned with the direction of $\mathbf{k}$ and uniform density $n_{b}$ passing through a
homogeneous electron plasma with density of electrons $n_{e}$. Therefore the partial contribution of the beam to the
total dielectric function of the beam--plasma system is given by Eq.~\eqref{eq:30}. For simplicity we will use throughout
the notation $u_{b\parallel} =u_{b}$ for the beam velocity. In the case the RF is off the chosen geometry corresponds
to the excitations of the TS unstable modes provided that the return plasma current is included in the dispersion
relations. Nevertheless, in the present study neglecting the return current we will adopt the terminology \textquotedblright
two--stream instability\textquotedblright. This should not be confusing as long as the velocity of the beam is parallel
to $\mathbf{k}$.

\section{Transversal polarization of the RF ($\mathbf{E}_{0}\perp \mathbf{k}$)}
\label{sec:4}

In this section we consider Eq.~\eqref{eq:25} for the harmonics of the electric field when the
polarization vector $\mathbf{E}_{0}$ of the laser field is perpendicular to $%
\mathbf{k}$ ($(\mathbf{k}\cdot \mathbf{E}_{0})=0$ and $\zeta _{\alpha }=0$).
Then from Eq.~\eqref{eq:21} it is seen that the nonvanishing components of
the conductivity tensor are $\sigma _{rj}^{(\ell ,\ell )}$, $\sigma
_{rj}^{(\ell ,\ell \pm 1)}$, $\sigma _{rj}^{(\ell ,\ell \pm 2)}\neq 0$ while
$\sigma _{rj}^{(\ell ,\ell \pm p)}=0$ at $p\geqslant 3$. Using this fact Eq.~%
\eqref{eq:25} for the electric field harmonics is represented as
\begin{eqnarray}
&&\left[ k^{2}\delta _{rj}-k_{r}k_{j}-\frac{(\omega +\ell \omega_{0})^{2}}{c^{2}}\Sigma _{rj}(\ell )\right]
E_{j}^{(\ell )}(0)  \nonumber \\
&& =D_{rj}^{+}(\ell ) E_{j}^{(\ell +1)}(0) +D_{rj}^{-}(\ell ) E_{j}^{(\ell -1)}(0) +R_{rj}^{+}(\ell )
E_{j}^{(\ell +2)}(0) \label{eq:33} \\
&& +R_{rj}^{-}(\ell ) E_{j}^{(\ell -2)}(0) +\frac{\omega ^{2}}{c^{2}}\delta _{\ell 0}\delta \varepsilon _{b,rj}
(\mathbf{k}, \omega ) E_{j}(0) ,  \nonumber
\end{eqnarray}%
where we have introduced the following notations $\mathbf{E}^{(n)}(\ell
)\equiv \mathbf{E}^{(n)}(\mathbf{k},\omega -\ell \omega _{0})$, $\mathbf{E}%
(\ell )\equiv \mathbf{E}(\mathbf{k},\omega -\ell \omega _{0})$, $\Sigma
_{rj}(\ell )\equiv \Sigma _{rj}(\mathbf{k},\omega +\ell \omega _{0})$, $%
D_{rj}^{\pm }(\ell )\equiv D_{rj}^{\pm }(\mathbf{k},\omega +\ell \omega
_{0}) $, $R_{rj}^{\pm }(\ell )\equiv R_{rj}^{\pm }(\mathbf{k},\omega +\ell
\omega _{0})$, $\delta \varepsilon _{b,rj}(\ell )\equiv \delta \varepsilon
_{b,rj}(\mathbf{k},\omega +\ell \omega _{0})$ and
\begin{eqnarray}
&&\Sigma _{rj}( \mathbf{k},\omega ) =\varepsilon _{rj}(
\mathbf{k},\omega ) +\frac{k^{2}\omega _{0}^{2}}{4\omega ^{2}}%
\sum_{\alpha }a_{\alpha r}a_{\alpha j}\left[ \delta \varepsilon _{\alpha
\parallel }( -1) +\delta \varepsilon _{\alpha \parallel }(
1) \right] ,  \label{eq:34} \\
&&D_{rj}^{\pm }( \mathbf{k},\omega ) =-\frac{\omega _{0}\omega }{%
2c^{2}}\sum_{\alpha }\left[ \frac{\omega }{\omega \pm \omega _{0}}%
k_{r}a_{\alpha j}\delta \varepsilon _{\alpha \parallel }( 0)
+k_{j}a_{\alpha r}\delta \varepsilon _{\alpha \parallel }( \pm 1) %
\right] ,  \label{eq:35} \\
&&R_{rj}^{\pm }( \mathbf{k},\omega ) =\frac{k^{2}\omega _{0}^{2}}{%
4c^{2}}\frac{\omega }{\omega \pm 2\omega _{0}}\sum_{\alpha }a_{\alpha
r}a_{\alpha j}\delta \varepsilon _{\alpha \parallel }( \pm 1) .
\label{eq:36}
\end{eqnarray}%
In the following we consider throughout an electron plasma neglecting the
dynamics of plasma ions. Thus we restrict ourself by the frequency domain
well above the ionic frequencies. We introduce the oscillation amplitude and
the quiver velocity of the electrons via relations $\mathbf{a}_{e}=-a\mathbf{%
e}$, $\mathbf{v}_{Ee}=-v_{E}\mathbf{e}$, $v_{E}=eE_{0}/m\omega _{0}$, $%
a=eE_{0}/m\omega _{0}^{2}$ ($-e$ is the electron charge), and $\mathbf{e}=%
\mathbf{E}_{0}/E_{0}$. So, the quantities $a$\ and $v_{E}$\ are positive by
definition.

Next, for exclusion of harmonics $E_{j}^{(\ell )}(0)$ in Eq.~\eqref{eq:33} the frequency $\omega $ is replaced
by $\omega -\ell \omega_{0}$ and using Eq.~\eqref{eq:24} we perform summation over $\ell $. This yields an
equation for the amplitude $E_{j}(0)$ of the electric field,
\begin{eqnarray}
&&\left[ k^{2}\delta _{rj}-k_{r}k_{j}-\frac{\omega ^{2}}{c^{2}}\Sigma_{rj}( 0) \right] E_{j}( 0)  \nonumber \\
&&=D_{rj}^{+}( 0) E_{j}( -1) +D_{rj}^{-}(0) E_{j}( 1) +R_{rj}^{+}( 0) E_{j}(-2)  \label{eq:37} \\
&&+R_{rj}^{-}( 0) E_{j}( 2) +\frac{\omega ^{2}}{c^{2}} \delta \varepsilon _{b,rj}( 0) E_{j}( 0) .  \nonumber
\end{eqnarray}%
The resulting equation represents an infinite system of linear equations for the quantities $E_{j}(\pm p)$ (with
$p=0,1,2,...$). The (infinite) determinant of this system determines the dispersion equation for the beam--plasma
system in the presence of the RF.

It follows from Eq.~\eqref{eq:37} that for the perturbations of which electric vector $\mathbf{E}(0)$ is polarized
perpendicular to the plane of the vectors $\mathbf{k}$ and $\mathbf{E}_{0}$, there is no instability. The dispersion
equation for these modes is given by $\mathcal{D}_{\perp }(k,\omega )=0$, where
\begin{equation}
\mathcal{D}_{\perp }( k,\omega ) =1+\frac{1}{k^{2}c^{2}}\left[\frac{\omega _{b}^{2}}{\gamma _{b}}-\omega ^{2}
\varepsilon _{\perp }(k,\omega ) \right] .
\label{eq:38}
\end{equation}%
This is simply the dispersion equation for the ordinary transverse electromagnetic modes propagating in an isotropic
plasma \cite{ale84,ich73} in the absence of the RF but modified due to the presence of the beam. The second term in
the right--hand side of Eq.~\eqref{eq:38} with minus sign is the partial contribution of the transverse dielectric
function of a cold beam to the total transverse dielectric function of the beam--plasma system. It is noteworthy that
this contribution depends on $\gamma _{b}^{-1}$ while the longitudinal contribution \eqref{eq:31} decays as $\gamma_{b}^{-3}$
with the beam relativistic factor.

To reveal the instability we consider therefore the case of a polarization wherein the electric vector $\mathbf{E}(0)$
of the perturbations lies in the plane containing the vectors $\mathbf{k}$ and $\mathbf{E}_{0}$. Introducing the
components of the electric field parallel ($E_{\parallel}$) and perpendicular ($\mathbf{E}_{\perp}$) to the wave vector
$\mathbf{k}$, for these modes from Eq.~\eqref{eq:37} we obtain
\begin{equation}
\mathcal{D}_{\parallel }( 0) E_{\parallel }( 0) =-\frac{kv_{E}}{2}\left[ \psi ( 1) +\psi ( -1) \right]
\delta \varepsilon _{\parallel }( 0) ,
\label{eq:39}
\end{equation}%
\begin{eqnarray}
&&\left\{ \mathcal{D}_{\perp }( 0) -\frac{v_{E}^{2}}{4c^{2}}\left[\delta \varepsilon _{\parallel }( 1) +\delta
\varepsilon_{\parallel }( -1) \right] \right\} \psi ( 0)  \nonumber \\
&&=\frac{v_{E}^{2}}{4c^{2}}\left[ \psi ( -2) \delta \varepsilon_{\parallel }( 1) +\psi ( 2) \delta
\varepsilon_{\parallel }( -1) \right]  \label{eq:40} \\
&&+\frac{v_{E}}{2kc^{2}}\left[ E_{\parallel }( -1) \delta \varepsilon _{\parallel }( 1) +E_{\parallel }(1)
\delta \varepsilon _{\parallel }( -1) \right] ,  \nonumber
\end{eqnarray}%
where $\mathcal{D}_{\parallel }(\ell )\equiv \mathcal{D}_{\parallel
}(k,\omega +\ell \omega _{0})$, $\mathcal{D}_{\perp }(\ell )\equiv \mathcal{D%
}_{\perp }(k,\omega +\ell \omega _{0})$, and $\psi (\mathbf{k},\omega )=(%
\mathbf{e}\cdot \mathbf{E}_{\perp }(\mathbf{k},\omega ))/\omega $ with $\psi
(\ell )\equiv \psi (\mathbf{k},\omega -\ell \omega _{0})$. Let us recall
that the function $\mathcal{D}_{\parallel }(0)$ given by Eq.~\eqref{eq:32}
is the total longitudinal dielectric function of the beam--plasma system in
the absence of the RF. In this case the transverse and longitudinal modes
are independent with the dispersion relations $\mathcal{D}_{\perp }(0)=0$
and $\mathcal{D}_{\parallel }(0)=0$, respectively. However, in the presence
of the laser radiation these modes are parametrically coupled according to
Eqs.~\eqref{eq:39} and \eqref{eq:40}. The longitudinal electric fields in
Eq.~\eqref{eq:40} can be excluded inserting the values $E_{\parallel }(-1)$
and $E_{\parallel }(1)$\ calculated by means of Eq.~\eqref{eq:39} into Eq.~%
\eqref{eq:40}. Then the given equation contains only the harmonics $\psi (0)$
and $\psi (\pm 2)$. Similarly, the transverse fields can be partially
excluded from Eq.~\eqref{eq:39} evaluating the harmonics $\psi (1)$ and $%
\psi (-1)$\ by means of Eq.~\eqref{eq:40}. In this case Eq.~\eqref{eq:39}
cannot be decoupled completely since it involves not only the longitudinal
electric fields but also the higher harmonics of the transverse fields. Also
it should be noted that the nonlinear response of the system is accompanied
by the magnetic field generation according to Eq.~\eqref{eq:15}. It
follows from this equation that the magnetic field is directed perpendicular
to the plane containing the vectors $\mathbf{k}$ and $\mathbf{E}_{0}$.

In principle the dispersion equations of the perturbations can be deduced
from Eqs.~\eqref{eq:39} and \eqref{eq:40} by solving these equations by
iteration to any order of accuracy. Taking, however, into account the
smallness of the parameter $v_{E}/c$, it suffices to restrict the analysis
of the system \eqref{eq:39} and \eqref{eq:40} to the harmonics $E_{\parallel
}(0)$, $E_{\parallel }(\pm 2)$ and $\psi (0)$, $\psi (\pm 2)$. In this case
the longitudinal and transversal modes are decoupled and we obtain the
following dispersion equations
\begin{eqnarray}
&&\mathcal{D}_{\parallel }( k,\omega ) +\frac{v_{E}^{2}}{4c^{2}}%
\left[ \frac{1}{\mathcal{D}_{\perp }( k,\omega -\omega _{0}) }+%
\frac{1}{\mathcal{D}_{\perp }( k,\omega +\omega _{0}) }\right]
\delta \varepsilon _{\parallel }^{2}( k,\omega ) =0 , \label{eq:41}  \\
&&\mathcal{D}_{\perp }( k,\omega ) =\frac{v_{E}^{2}}{4c^{2}}%
\sum_{\nu =\pm }\frac{\delta \varepsilon _{\parallel }( k,\omega +\nu
\omega _{0}) \varepsilon _{b\parallel }( k,\omega +\nu \omega
_{0}) }{\mathcal{D}_{\parallel }( k,\omega +\nu \omega_{0}) } \label{eq:42}
\end{eqnarray}%
for longitudinal and transversal modes, respectively. Here $\varepsilon_{b\parallel }(k,\omega )=1+\delta
\varepsilon _{b\parallel }(k,\omega )$.

\subsection{Longitudinal modes}
\label{sec:4.1}

Let us now investigate the dispersion equations \eqref{eq:41} and \eqref{eq:42} in detail within fluid model (or
cold plasma approximation) when the partial dielectric functions are given by
\begin{equation}
\delta \varepsilon _{\perp }(k,\omega ) =\delta \varepsilon_{\parallel }(k,\omega ) \equiv \delta
\varepsilon (\omega ) =-\frac{\omega _{p}^{2}}{\omega ^{2}} .
\label{eq:cold}
\end{equation}%
Here $\omega _{p}^{2}=4\pi n_{e}e^{2}/m$ is the plasma frequency. Thus we
consider only the high--frequency modes assuming that $\vert \omega\vert \gg kv_{\mathrm{th}}$, where $v_{\mathrm{th}}$ is the thermal
velocity of the electrons. We look for the solutions of the dispersion
equations in the form $\omega =\omega _{r}+i\gamma $, where $\omega _{r}$\
is the (real) frequency and $\gamma $ is the damping rate (when $\gamma <0$)
or the growth rate (when $\gamma >0$) of the modes, respectively. In the
absence of the laser field ($v_{E}=0$) the transverse modes are stable and
their frequency is determined by
\begin{equation}
\omega _{\perp }^{2}(k ) =k^{2}c^{2}+\frac{\omega _{b}^{2}}{\gamma _{b}}+\omega _{p}^{2} .
\label{eq:tr}
\end{equation}%
It is seen that the frequency of the ordinary transverse modes is modified
by the charged particles beam effectively increasing the total plasma
frequency of the beam--plasma system. Also it should be noted that the
contribution of the beam to the dispersion relation of the transverse modes
is $\sim \omega _{b}^{2}/\gamma _{b}$ while for the longitudinal modes it is
given by $\sim \omega _{b}^{2}/\gamma _{b}^{3}$ (see, e.g., Eq.~\eqref{eq:31}).
This is a consequence of the anisotropy of the effective electron mass
with respect to the driving force acting either in longitudinal or
transversal directions. The stability of the mode \eqref{eq:tr} can be
easily understood taking into account the fact that the electric field
vector in this mode is perpendicular to the beam and hence the work
performed by this field on the beam particles is zero.

The longitudinal two--stream modes are unstable in a long wavelength regime
(see, e.g., Ref.~\cite{mik74}) with $0\leqslant k\leqslant k_{c}\equiv
\omega _{c}/u_{b}$, where $\omega _{c}=\omega _{p}(1+\xi ^{1/3}/\gamma
_{b})^{3/2}$ and $\xi =\omega _{b}^{2}/\omega _{p}^{2}$. Note that in
practice $\xi \ll 1$ and $\omega _{c}\simeq \omega _{p}$. Assuming that $%
k\ll k_{c}$ the growth rate and the frequency of the two--stream modes read
(cf. with Eqs.~\eqref{eq:cp4} and \eqref{eq:cp5})
\begin{equation}
\gamma ^{\mathrm{TS}}(k) \simeq ku_{b}\frac{(\xi /\gamma _{b}^{3}) ^{1/2}}{1+\xi /\gamma _{b}^{3}} ,
\qquad
\omega _{\parallel }^{\mathrm{TS}}(k) \simeq \frac{ku_{b}}{1+\xi /\gamma _{b}^{3}},
\label{eq:TS}
\end{equation}%
respectively. It is seen that the real frequency of the TS modes $\simeq ku_{b}$, i.e., it is a frequency--locked
oscillation, the frequency depending only the \v{C}herenkov--type term and not on the natural frequency of the
oscillations ($\sim \omega _{p}$). The maximal value of $\gamma ^{\mathrm{TS}}(k)$ is achieved at $k\lesssim k_{c}$
\cite{mik74} and is given by
\begin{equation}
\frac{\gamma _{\max }^{\mathrm{TS}}}{\omega _{p}}\simeq \frac{\sqrt{3}}{%
2^{4/3}}\frac{\xi ^{1/3}}{\gamma _{b}}.  \label{eq:maxTS}
\end{equation}

In the presence of the laser field ($v_{E}\neq 0$) the high--frequency
transversal and the low--frequency longitudinal modes, Eqs.~\eqref{eq:tr} and %
\eqref{eq:TS}, \eqref{eq:maxTS}, respectively, of the beam--plasma system are
parametrically coupled according to Eqs.~\eqref{eq:41} and \eqref{eq:42}.
These equations can be satisfied only when one of the ordinary dispersion
functions $\mathcal{D}_{\parallel }(k,\omega )$ or $\mathcal{D}_{\perp
}(k,\omega )$, becomes nearly equal to zero. This is not, however,
sufficient to cause parametric TS instability, which occurs when at
least two of the zeros of the dispersion functions merge, as in the case of
the standard TS instability \cite{mik74,ale84}. In the case of the
longitudinal waves there are three such situations: (i) \v{C}herenkov--type
coupling when $\omega _{r}\simeq ku_{b}$; (ii) $\mathcal{D}_{\parallel
}(k,\omega )\simeq 0$ and $\mathcal{D}_{\perp }(k,\omega \pm \omega
_{0})\simeq 0$; (iii) $\mathcal{D}_{\perp }(k,\omega +\omega _{0})\simeq 0$
and $\mathcal{D}_{\perp }(k,\omega -\omega _{0})\simeq 0$. The cases (i) and
(ii) correspond to the resonant coupling, and the case (iii) to the
nonresonant coupling.

Consider the situation (i) which corresponds to the low--frequency ($\omega
\sim \omega _{\parallel }^{\mathrm{TS}}(k)$), long--wavelength excitations.
Close to the \v{C}herenkov resonance, $\omega \simeq ku_{b}$, the most
important term in Eq.~\eqref{eq:41} is involved in $\delta \varepsilon
_{b\parallel }(\mathbf{k},\omega )$ (Eq.~\eqref{eq:31}). Consequently we
look for the solution of the dispersion equation \eqref{eq:41} in the form $%
\omega =ku_{b}+\omega _{1}$, where $\vert \omega _{1}\vert \ll
ku_{b}$. In this case the instability occurs at $ku_{b}\lesssim \omega
_{s}\sim \omega _{p}$ with the growth rate ($\omega _{1}=i\gamma $) (cf.
with Eq.~\eqref{eq:TS})
\begin{equation}
\gamma (k)\simeq \left( \frac{\xi }{\gamma _{b}^{3}}\right) ^{1/2}\frac{%
ku_{b}}{\sqrt{F(ku_{b})}} ,
\label{eq:a1}
\end{equation}%
where $\omega _{s}$ is the zero of the function $F(\omega )$, $F(\omega_{s})=0$, and
\begin{equation}
F(\omega ) =1-\frac{\omega ^{2}}{\omega _{p}^{2}}-\frac{v_{E}^{2}%
}{4c^{2}}\frac{\omega _{p}^{2}}{\omega ^{2}}\left[ \frac{1}{\mathcal{D}%
_{\perp }(k,\omega -\omega _{0} )}+\frac{1}{\mathcal{D}_{\perp
}(k,\omega +\omega _{0} )}\right] .
\label{eq:a2}
\end{equation}%
It is noteworthy that the root $k_{0}$ of the transversal dispersion function, $\mathcal{D}_{\perp}(k,
\omega -\omega_{0}) \simeq 0$ with $\omega =ku_{b}$, at $\omega_{0}\lesssim 2\omega_{p}$ may lie in the
domain $ku_{b}\lesssim \omega_{s}$ and the growth rate~\eqref{eq:a1} changes the slope close to this root.
At $k \gtrsim k_{0}$ the function $F(\omega )$ becomes negative and the relation~\eqref{eq:a1} is violated.
However, at $\omega_{0}\gtrsim 2\omega_{p}$ the formula~\eqref{eq:a1} remains valid in the whole domain of
the instability. Also the maximal growth rate is achieved at $ku_{b}\simeq \omega _{s}$. Equation~\eqref{eq:a1}
is clearly invalid in this case and more rigorous treatment of the dispersion equation yields
\begin{equation}
\gamma _{\max }\simeq \frac{\sqrt{3}}{2^{4/3}}\left[ \frac{\xi }{\gamma
_{b}^{3}}\frac{2\omega _{s}^{2}}{\left\vert F^{\prime }(\omega_{s}) \right\vert }\right] ^{1/3} .
\label{eq:a3}
\end{equation}%
Here the prime indicates the derivative of the function with respect to the argument.

We have considered above the low--frequency and long wavelength regime when
the dispersion properties of the system are strongly determined by the beam
characteristics (density and the energy). Next we consider high--frequency ($%
\omega >\omega _{\parallel }^{\mathrm{TS}}(k)$), short wavelength regime
with $ku_{b}\gtrsim \omega _{s}$. This case corresponds to the situations
(ii) and (iii) introduced above. Assuming low density beam with $n_{b}\ll n_{e}$, we note that the role of the beam in the
dispersion properties of the system becomes less pronounced in this
high--frequency regime, and as a first approximation the terms proportional
to $\omega _{b}^{2}$ can be neglected in Eq.~\eqref{eq:41}. This is a regime
of purely parametric excitations in a plasma.

The resonant coupling (ii) occurs when the following resonance condition is satisfied:
\begin{equation}
\omega _{0}\simeq \omega _{p}+\omega _{\perp }(k) .
\label{eq:res}
\end{equation}%
In either case, $\vert \omega \vert$ is assumed to be much
smaller than $\omega _{0}$\ and $\omega _{\perp }(k)$. We can them make the
resonance approximation for the dispersion function $\mathcal{D}_{\perp
}(k,\omega \pm \omega _{0})$ to obtain
\begin{equation}
\mathcal{D}_{\perp }(k,\omega \pm \omega _{0}) \simeq \mp \frac{1}{k^{2}c^{2}}(2\omega _{0}-\delta )(\omega \pm \delta ) ,
\label{eq:a4}
\end{equation}%
where $\delta (k)=\omega _{0}-\omega _{\perp }(k)$ is the mismatch of the
laser frequency from the frequency of the natural transversal mode.
Substituting Eq.~\eqref{eq:a4} into \eqref{eq:41} and neglecting the
contribution of the beam in the dispersion relation of the left hand side of
Eq.~\eqref{eq:41} yields a cubic equation for $\omega ^{2}$,
\begin{equation}
\omega ^{2} (\omega ^{2}-\omega _{p}^{2}) (\omega^{2}-\delta ^{2} ) +\frac{k^{2}v_{E}^{2}}{4}
\frac{\omega _{p}^{4}\delta}{\omega _{0}-\delta /2}=0
\label{eq:a5}
\end{equation}%
which can be solved analytically. However, we restrict ourself to the
simple and qualitative solutions of the dispersion equation and more
rigorous numerical calculations will be presented in Sec.~\ref{sec:6}.

As mentioned above there are two types of solutions of the high--frequency ($%
\omega >\omega _{\parallel }^{\mathrm{TS}}(k)$) dispersion equation %
\eqref{eq:a5}. First we consider the resonant solution assuming that $\delta
\simeq \omega _{p}>0$. Note that this type of instability resembles the
resonant decay instability. The sum frequency of the excited modes is
exactly equal to the laser frequency $\omega _{0}$ (see Eq.~\eqref{eq:res}).
Introducing the frequency mismatch $\Delta (k)=\delta (k)-\omega _{p}$ ($%
\Delta \ll \omega _{p}$, $\delta $) the dispersion equation \eqref{eq:a5}
for the resonant growth rate yields
\begin{equation}
\gamma _{r}(k)=\frac{1}{2}\sqrt{\epsilon ^{2}k^{2}c^{2} -\Delta ^{2}(k)} ,
\label{eq:a6}
\end{equation}%
where $\epsilon =(v_{E}/2c)(\tau -1/2)^{-1/2}$, $\tau =\omega _{0}/\omega
_{p}$. The resonant unstable mode exists only at $\tau \gtrsim 2$. The
maximal growth rate $\gamma _{r,\max }$ is achieved at $k=k_{r,\max }$ with
\begin{eqnarray}
&&k_{r,\max } =\frac{\omega _{p}}{c}\frac{\sqrt{(\tau -2+\epsilon
^{2}) (\tau -\epsilon ^{2} )}}{1-\epsilon ^{2}} , \label{eq:a7} \\
&&\gamma _{r,\max } =\frac{\omega _{p}}{2}\epsilon \sqrt{\frac{\tau (
\tau -2) +\epsilon ^{2}}{1-\epsilon ^{2}}} .  \label{eq:a8}
\end{eqnarray}

Now in this high--frequency domain the characteristics of the instability are only weakly sensitive to the
beam density and energy ($\gamma _{b}$) but essentially depend on the laser intensity and the frequency. It
is seen that the maximal growth rate is scaled as (at $\tau \gg 1$) $\gamma _{r,\max}\sim [I_{L}(\omega%
_{p}/\omega _{0})]^{1/2}$, where $I_{L}$ is the RF intensity. It is also noteworthy that at
$\tau \gg 1$ the position $k_{r,\max }$ of the maximum of the resonant growth rate is independent on the laser
intensity ($k_{r,\max}\simeq \omega _{0}/c$). Equations~\eqref{eq:a6}--\eqref{eq:a8} can be compared with the
growth rate of the ordinary two--stream instability, Eq.~\eqref{eq:maxTS}. Assuming, for simplicity,
high--frequency laser field ($\omega _{0}\gg \omega _{p}$) one obtains that at $v_{E}/c>\sqrt{3/\tau }(4\xi%
/\gamma _{b}^{3})^{1/3}$ the growth rate $\gamma _{r,\max }$ exceeds $\gamma _{\max }^{\mathrm{TS}}$. Note that
the last inequality is easily fulfilled for the REB.

In the nonresonant case (iii) assuming that $\vert\omega \vert \ll \omega _{p}$ from Eq.~\eqref{eq:a5}
one obtains a quadratic equation for $\omega ^{2}$. A simple analysis of this equation shows that in this
case the instability occurs at $-\delta _{m}\leqslant \delta \leqslant\omega _{0}-\omega _{p}$, where
$\delta _{m}=\omega _{p}[2\epsilon_{1}(kc/\omega _{p})]^{2/3}$ with $\epsilon _{1}=(v_{E}/2c)(\tau -\delta
/2\omega _{p})^{-1/2}$. Two distinct branches of the instability should be considered separately. At the
positive frequency mismatch, $0\leqslant\delta \leqslant \omega _{0}-\omega _{p}$, (or $k\leqslant k_{2}
\equiv (\omega_{p}/c)(\tau ^{2}-1)^{1/2}$) the instability is almost aperiodic (i.e. $\omega_{r}\simeq 0$)
with the growth rate
\begin{equation}
\gamma =\frac{\delta ^{1/4}}{\sqrt{2}}\left( \sqrt{\delta ^{3}+\delta
_{m}^{3}}-\delta ^{3/2}\right) ^{1/2} ,
\label{eq:a9}
\end{equation}%
while at the negative values with $-\delta _{m}\leqslant \delta \leqslant 0$ ($k_{2} \leqslant k\leqslant
k_{1}$, where $\delta =-\delta _{m}$ at $k=k_{1}$) the instability is periodic (i.e. $\omega_{r}\neq 0$)
and the growth rate becomes
\begin{equation}
\gamma =\frac{\left\vert \delta \right\vert ^{1/4}}{2}\left( \delta
_{m}^{3/2}-\left\vert \delta \right\vert ^{3/2}\right) ^{1/2} .
\label{eq:a10}
\end{equation}%
The real frequency of the unstable mode~\eqref{eq:a10} is obtained by changing the minus sign in this formula
by plus sign. We note that the latter nonresonant case with $\delta <0$ resembles the oscillating two--stream
instability.

Thus summarizing this section we emphasize that in the spectrum of the longitudinal unstable modes there are
basically three domains with strictly different properties. The "long wavelength" domain with $k\lesssim \omega
_{p}/u_{b}$ (we denote this parameter regime as Domain I) is basically determined by the beam density and the
energy ($\gamma _{b}$) and corresponds to the TS instability (see, e.g., Eqs.~\eqref{eq:maxTS} and
\eqref{eq:a1}). The intermediate (Domain II, $k\leqslant k_{2}$) and the short wavelength (Domain III,
$k_{2}\leqslant k\leqslant k_{1}$) regimes mainly depend on the laser intensity and are only weakly sensitive
to the beam parameters. In II and III the growth rates can be approximated by Eqs.~\eqref{eq:a9} and \eqref{eq:a10},
respectively. In addition in Domain II at $\omega _{0}\gtrsim 2\omega _{p}$ it is possible to witness a resonant
excitation of the unstable modes with maximal growth rate \eqref{eq:a8} which resembles the resonant decay instability.
Finally, the Domain I may merge to Domain II at $k\simeq \omega _{p}/u_{b}$ while the Domain II merges to
Domain III at zero frequency mismatch, $\delta =0$ ($k=k_{2}$).

\subsection{Transversal modes}
\label{sec:4.2}

In this section we turn to the investigation of the unstable transversal
modes generated in the beam--plasma system. Our starting point is the
dispersion equation \eqref{eq:42} for these modes. As in Sec.~\ref{sec:4.1}
we adopt here a cold--fluid approximation when the partial dielectric
functions of the beam and the plasma are given by Eqs.~\eqref{eq:31} and %
\eqref{eq:cold}, respectively. An inspection of the dispersion equation %
\eqref{eq:42} shows that there is only one resonant coupling between
different modes. This is the situation when $\mathcal{D}_{\perp }(k,\omega
)\simeq 0$ and $\mathcal{D}_{\parallel }(k,\omega -\omega _{0})\simeq 0$.
This system of equations require that $\omega \simeq \omega _{0}+\omega
_{s}(k)+\Delta \omega (k)$ and $\omega _{\perp }(k)\simeq \omega _{0}+\omega
_{s}(k)$,\ where $\Delta \omega \ll \omega _{0}+\omega _{s}$, $\omega
_{\perp }(k)$\ is the frequency of the ordinary transversal modes, Eq.~%
\eqref{eq:tr}, and $\omega _{s}$ is the real root of the ordinary dispersion
equation, $\mathcal{D}_{\parallel }(k,\omega _{s})=0$, for the longitudinal
modes. Note that the resonant condition $\omega _{\perp }\simeq \omega
_{0}+\omega _{s}$\ cannot be satisfied in the domain where the two--stream
instability occurs, where $\omega _{s}(k)=\omega _{\parallel }^{\mathrm{TS}%
}(k)+i\gamma ^{\mathrm{TS}}(k)$ (see, e.g., Eqs.~\eqref{eq:TS} and
\eqref{eq:maxTS}) is a complex quantity. Therefore it is expected that the
resonance occurs at short wavelengths ($k\gtrsim k_{c}$) and at
high--frequencies ($\omega \gtrsim \omega _{\parallel }^{\mathrm{TS}}(k)$).
Inserting $\omega \simeq \omega _{0}+\omega _{s}+\Delta \omega $ and $\omega
_{\perp }\simeq \omega _{0}+\omega _{s}$ into Eq.~\eqref{eq:42} and
neglecting a small term depending on the frequency $\omega +\omega _{0}$ in
the right--hand side of this equation, for the maximal growth rate we obtain
\begin{equation}
\frac{\gamma _{r,\max }}{\omega _{p}}\simeq \frac{v_{E}}{4c}\sqrt{-\frac{%
2\omega _{p}^{2}}{\omega _{\perp }\omega _{s}^{4}}\frac{k^{2}c^{2}}{\frac{%
\partial }{\partial \omega }\mathcal{D}_{\parallel }(k,\omega _{s} )}} .
\label{eq:x1}
\end{equation}%
In addition, the resonant instability occurs only for a negative derivative,
$\frac{\partial }{\partial \omega }\mathcal{D}_{\parallel }(k,\omega _{s})<0$,
of the longitudinal dispersion function. An analysis shows that there is
only one real root $\omega _{s}=\omega _{r1}^{-}$\ of the equation $\mathcal{%
D}_{\parallel }(k,\omega _{s})=0$ which satisfies this condition (see
Appendix~\ref{sec:app0} for details). This root is negative and is
represented here as $\omega _{r1}^{-}(k)=-\omega _{p}g(k)$, where
the function $g(k)$ is positive and decreases monotonically from $%
(1+\xi /\gamma _{b}^{3})^{1/2}$ (at $k\simeq 0$) to $1$ (at $k\gg k_{c}$),
see Eqs.~\eqref{eq:cp6} and \eqref{eq:cp1}, respectively.

The maximal growth rate \eqref{eq:x1} is reached at $k=k_{\max }$\ which is
determined by the resonant condition above, $\omega _{\perp }(k)=\omega
_{0}-\omega _{p}g(k)$. Assuming short wavelengths ($k\gtrsim k_{c}$%
) this equation can be solved iteratively. The result reads as
\begin{equation}
\frac{k_{\max }^{2}c^{2}}{\omega _{p}^{2}} \simeq \left[ \tau -g(\kappa )\right] ^{2}-\frac{\xi }{\gamma _{b}}-1 ,
\label{eq:x2}
\end{equation}%
where $\kappa ^{2}c^{2}=\omega _{p}^{2}[\tau (\tau -2)-\xi /\gamma _{b}]$.
Note that the resonant instability occurs only at high--frequencies of the
RF, $\tau \gtrsim \tau _{c}\equiv g(\kappa )+(1+\xi /\gamma
_{b})^{1/2}$ (for a low--density electron beam this condition is roughly
equivalent to $\omega _{0}\gtrsim 2\omega _{p}$).

Substituting Eq.~\eqref{eq:x2} into Eq.~\eqref{eq:x1} we arrive at
\begin{equation}
\frac{\gamma _{r,\max }}{\omega _{p}}\simeq \frac{v_{E}}{4c}\sqrt{\frac{\left[\tau -g(\kappa )\right]^{2}
-\xi /\gamma _{b}-1}{g(\kappa ) \left[ \tau -g(\kappa )\right] }\frac{1}{1+(\xi /\gamma _{b}^{3})H(\kappa )}}
\label{eq:x3}
\end{equation}%
with
\begin{equation}
H(\kappa ) =\left[ \frac{g(\kappa )}{\kappa u_{b}/\omega _{p}+g(\kappa )}\right] ^{3} .
\label{eq:x4}
\end{equation}

The maximal growth rate for the resonant instability is strongly simplified
for a very low density ($n_{b}\ll n_{e}$) or for an ultrarelativistic
electron beam ($\gamma _{b}\gg 1$). In this case the instability occurs at $%
\tau \gtrsim 2$ and $k_{\max }c=\omega _{p}[\tau (\tau -2)]^{1/2}$. The
growth rate $\gamma _{r,\max }$ is then simply given by
\begin{equation}
\frac{\gamma _{r,\max }}{\omega _{p}}\simeq \frac{v_{E}}{4c}\sqrt{\frac{\tau
\left( \tau -2\right) }{\tau -1}} ,
\label{eq:x5}
\end{equation}%
and is completely independent on the beam parameters (this is a regime of a
purely parametric instability). Note that similar to Eq.~\eqref{eq:a8} for
the longitudinal resonant unstable mode the resonant growth rate %
\eqref{eq:x5} at $\omega _{0}\gg \omega _{p}$ is scaled as $\gamma _{r,\max
}\sim [I_{L}(\omega _{p}/\omega _{0})]^{1/2}$. Thus the growth rates
of the resonant longitudinal and transversal unstable modes may be of the
same order.

Besides the resonant mode there also exist two nonresonant transversal
modes. The dispersion equations for these modes follow from Eq.~\eqref{eq:42}
and are given by $\mathcal{D}_{\parallel }(k,\omega \pm \omega _{0})\simeq 0$
which implies that $\omega =\mp \omega _{0}+\omega _{s}+\Delta \omega _{\pm }
$ (with $|\omega _{\pm }|\ll |\mp \omega _{0}+\omega _{s}|$). Here $\omega
_{s}=\omega _{\parallel }^{\mathrm{TS}}+i\gamma ^{\mathrm{TS}}$ (see, e.g.,
Eqs.~\eqref{eq:TS} and \eqref{eq:cp4}, \eqref{eq:cp5}) is the solution of
the longitudinal dispersion equation in the domain of the two--stream
instability which occurs at the low--frequencies ($\omega \simeq \omega
_{\parallel }^{\mathrm{TS}}(k)$) and at the long wavelengths ($k\leqslant
k_{c}$).

The quantity $\Delta \omega _{\pm }$ can be roughly estimated using the
dispersion equation \eqref{eq:42}. First, in the leading order we represent
the dispersion function in the form $\mathcal{D}_{\parallel }(k,\omega \pm
\omega _{0})\simeq \Delta \omega _{\pm }\frac{\partial }{\partial \omega }%
\mathcal{D}_{\parallel }(k,\omega _{s})$. Second, the derivative of the
dispersion function is estimated employing Eqs.~\eqref{eq:cp4} and %
\eqref{eq:cp5}. Assuming a low--density ($n_{b}\ll n_{e}$) or an
ultrarelativistic ($\gamma _{b}\gg 1$) electron beam this quantity in the
leading order of the dimensionless parameter $\xi /\gamma _{b}^{3}$ is
represented as $\frac{\partial }{\partial \omega }\mathcal{D}_{\parallel
}(k,\omega _{s})\simeq (2i/\gamma ^{\mathrm{TS}})(\omega _{p}/\omega
_{\parallel }^{\mathrm{TS}})^{2}$. Then inserting these results into
dispersion equation \eqref{eq:42} in the leading order of $\xi /\gamma
_{b}^{3}$ one finally arrives at
\begin{equation}
\Delta \omega _{\pm }\simeq -i\frac{k^{2}v_{E}^{2}}{8}\frac{\omega
_{p}^{2}\gamma ^{\mathrm{TS}}}{(\omega _{\parallel }^{\mathrm{TS}%
})^{2}[(\omega _{0}\mp \omega _{\parallel }^{\mathrm{TS}})^{2}-\omega
_{\perp }^{2}]}.  \label{eq:x6}
\end{equation}%
The real frequencies and the growth rates of the nonresonant modes are
simply given by $\omega _{r}=\mp \omega _{0}+\omega _{\parallel }^{\mathrm{TS%
}}$ and\ $\gamma =\gamma ^{\mathrm{TS}}+\mathrm{Im}[\Delta \omega _{\pm }]$,
respectively. It should be emphasized that the frequency shift \eqref{eq:x6}
is valid when $\mathrm{Im}[\Delta \omega _{\pm }]\ll \gamma ^{\mathrm{TS}}$.
Moreover, since the nonresonant unstable modes appear in the domain of the
two--stream instability with $k\leqslant k_{c}$\ the frequencies $\omega
_{\parallel }^{\mathrm{TS}}$ and $\omega _{\perp }$\ in the brackets of the
denominator of Eq.~\eqref{eq:x6} can be neglected and hence $\Delta \omega
_{+}\simeq \Delta \omega _{-}$. The expression \eqref{eq:x6} can be further
simplified recalling that $\omega _{\parallel }^{\mathrm{TS}}\simeq ku_{b}$\
and the ratio $\gamma ^{\mathrm{TS}}/\omega _{\parallel }^{\mathrm{TS}%
}\simeq (\xi /\gamma _{b}^{3})^{1/2}$ is almost independent on $k$ (see,
e.g., Eqs.~\eqref{eq:cp4} and \eqref{eq:cp5}). Therefore the growth rate of
the nonresonant modes is given by
\begin{equation}
\gamma (k) \simeq \gamma ^{\mathrm{TS}}(k) \left[1- \frac{1}{8}\left( \frac{v_{E}}{u_{b}}\right)^{2}
\left( \frac{\omega _{p}}{\omega _{0}}\right) ^{2}\right] .
\label{eq:x7}
\end{equation}%
It is noteworthy that this growth rate is proportional to the growth rate $%
\gamma ^{\mathrm{TS}}(k)$ of the standard two--stream instability and only
the factor in the brackets depends on the laser intensity. Thus $\gamma (k)$
is weakly sensitive to the laser intensity (because in the present
approximation $v_{E}\ll u_{b}$ and $\omega _{0}>\omega _{p}$) and is mainly
determined by the beam--plasma interaction. In addition, the nonresonant
unstable modes do not disappear with decreasing the laser intensity as it
occurs for the resonant ones.

\section{Longitudinal polarization of the RF ($\mathbf{E}_{0}\parallel \mathbf{k}$)}
\label{sec:5}

With the theoretical formalism presented in Secs.~\ref{sec:2} and \ref{sec:3}, we now take up another configuration
of the laser polarization. In the following we study in detail the parametric two--stream instabilities in the laser
irradiated plasma when the polarization vector of the laser field is parallel to the wave vector $\mathbf{k}$ of the
excitations ($\mathbf{E}_{0}\parallel \mathbf{k}$) assuming again that the beam is directed in the direction of
$\mathbf{k}$. It is expected that the beam--plasma and the laser--plasma unstable modes are strongly coupled in this
regime compared to the transversal configuration ($\mathbf{E}_{0}\perp \mathbf{k}$) since the electrons are effectively
driven by a laser radiation in the direction of $\mathbf{k}$ ($\parallel \mathbf{u}_{b}$) in this case.

Our starting point is the general equation~\eqref{eq:25} for the harmonics which for the configuration $\mathbf{E}_{0}%
\parallel \mathbf{k}\parallel \mathbf{u}_{b}$ is decoupled into two independent equations for the longitudinal ($E_{\parallel }$)
and transversal ($\mathbf{E}_{\perp }$) electric fields, respectively,
\begin{equation}
E_{\parallel }^{(n)}( 0) +\sum_{\ell ,s=-\infty }^{\infty} E_{\parallel }^{(\ell )}( 0) J_{\ell -s}( \zeta )
J_{n-s}( \zeta ) \delta \varepsilon _{\parallel }( s) =-\delta _{n0}E_{\parallel }( 0) \delta
\varepsilon _{b\parallel}( 0) ,
\label{eq:43}
\end{equation}%
\begin{eqnarray}
&&\mathbf{E}_{\perp }^{(n)}( 0) =\frac{\omega +n\omega _{0}}{%
k^{2}c^{2}-( \omega +n\omega _{0}) ^{2}}\sum_{\ell ,s=-\infty
}^{\infty }\mathbf{E}_{\perp }^{(\ell )}( 0) \frac{( \omega
+s\omega _{0}) ^{2}}{\omega +\ell \omega _{0}}J_{\ell -s}( \zeta
) J_{n-s}( \zeta ) \delta \varepsilon _{\perp }(s)  \label{eq:44} \\
&&-\delta _{n0}\frac{1}{k^{2}c^{2}-\omega ^{2}}\frac{\omega _{b}^{2}}{\gamma_{b}}\mathbf{E}_{\perp }( 0) .  \nonumber
\end{eqnarray}%
Here $\zeta \equiv a(\mathbf{k}\cdot \mathbf{e})=\pm ka$ and the other notations have been introduced in Secs.~\ref{sec:2}
and \ref{sec:3}. As in the preceding Sections we consider throughout an electron plasma neglecting the dynamics of plasma
ions.

To exclude the electric field harmonics $E_{\parallel }^{(n)}$ in Eq.~\eqref{eq:43} we multiply both sides of this
equation by $J_{n-p}(\zeta )J_{\ell -p}(\zeta )$ and perform a summation over $n$ and $p$. As a result we arrive at
\begin{equation}
E_{\parallel }^{(n)}( 0) =-E_{\parallel }( 0) \delta \varepsilon _{b\parallel }( 0) \sum_{s=-\infty }^{\infty }
\frac{J_{s}( \zeta ) J_{n+s}( \zeta ) }{1+\delta \varepsilon_{\parallel }( -s) } .
\label{eq:45}
\end{equation}%
Thus the harmonics $E_{\parallel }^{(n)}$ are completely expressed by the total field $E_{\parallel }$. Also in deriving
Eq.~\eqref{eq:45} we have used the summation formula for the Bessel functions \cite{gra80}
\begin{equation}
\sum_{s=-\infty }^{\infty }J_{s-\ell }( \zeta ) J_{s-n}(\zeta ) =\delta _{\ell n} .
\label{eq:sum}
\end{equation}

Next, in Eq.~\eqref{eq:45} the frequency $\omega $\ is replaced by $\omega -n\omega _{0}$ and using Eq.~\eqref{eq:24} we
perform a summation over $n$. This yields an equation for the amplitude $E_{\parallel }(0)$ of the electric field
\begin{equation}
\left[ 1+\delta \varepsilon _{b\parallel }(0) \right] E_{\parallel }(0) =\sum_{\ell =-\infty }^{\infty }
E_{\parallel}(\ell ) \Psi _{\ell }(\zeta ) \delta \varepsilon_{b\parallel }(-\ell ) ,
\label{eq:47}
\end{equation}%
where
\begin{equation}
\Psi _{\ell }(\zeta ) =\sum_{s=-\infty }^{\infty }\frac{\delta \varepsilon _{\parallel }(-s)}{1+\delta%
\varepsilon _{\parallel}(-s)}J_{s-\ell }(\zeta ) J_{s}(\zeta ) .
\label{eq:psi}
\end{equation}%
It is noteworthy that in the $\mathbf{E}_{0}\parallel \mathbf{k}$ geometry the system of equations \eqref{eq:43}
and \eqref{eq:44} involves all (infinite number) the harmonics of the electric field whereas in the transversal case,
$\mathbf{E}_{0}\perp \mathbf{k}$, considered in Sec.~\ref{sec:4} each harmonic $E^{(n)}(0)$ with given frequency $\omega$
connects only to the nearest neighbors $E^{(n\pm 1)}(0)$ and $E^{(n\pm 2)}(0)$. These features are the peculiarities
of the specific laser polarization directed parallel or perpendicular to the wave vector $\mathbf{k}$.

Similarly, it is possible to exclude the harmonics $\mathbf{E}_{\perp}^{(n)} $ of the transversal electric field from
Eq.~\eqref{eq:44} and derive an equation for the electric field $\mathbf{E}_{\perp }$,
\begin{equation}
\mathbf{E}_{\perp }(0)\left( k^{2}c^{2}+\frac{\omega _{b}^{2}}{\gamma _{b}}%
-\omega ^{2}\right) =\omega \sum_{\ell ,s=-\infty }^{\infty }\mathbf{E}%
_{\perp }(-\ell )\frac{(\omega +s\omega _{0} )^{2}}{\omega +\ell
\omega _{0}}J_{\ell -s}(\zeta )J_{-s}(\zeta )\delta \varepsilon _{\perp }(s).
\label{eq:48}
\end{equation}

First, we note that at vanishing laser intensity from Eqs.~\eqref{eq:47}--\eqref{eq:48} we recover the standard dispersion
equations for the longitudinal, $\mathcal{D}_{\parallel }(k,\omega )=0$, and transversal, $\mathcal{D}_{\perp }(k,\omega )=0$,
waves given by Eqs.~\eqref{eq:32} and \eqref{eq:38}, respectively. Second, within cold--fluid approximation (the dielectric
functions are determined by Eq.~\eqref{eq:cold}) using the summation formula \eqref{eq:sum} for the transversal modes we
arrive at the same dispersion relation as in Eq.~\eqref{eq:tr}. Therefore within this approximation the laser field has
no influence on the dispersion properties of the transversal modes which are stable in this case. Third, in contrast to
the longitudinal modes \eqref{eq:47} the transversal ones are only weakly sensitive to the beam parameters (see left--hand
side of Eq.~\eqref{eq:48}) and the instability of these modes is parametric in nature. These instabilities have been studied
previously in Ref.~\cite{gor66}. Consequently, in the following we consider throughout only the dynamics of the longitudinal
modes $E_{\parallel }$.

As was mentioned in the previous sections the dispersion equation of the perturbations can be deduced, in principle, from
Eq.~\eqref{eq:47} by solving this system of equations by iteration to any order of accuracy. However, to gain more insight
we consider the long wavelength limit of Eq.~\eqref{eq:47} when the parameter $|\zeta |=ka$ is small, which suffices to
restrict the analysis of the system \eqref{eq:47} to the harmonics $E_{\parallel }(0)$ and $E_{\parallel }(\pm 1)$. In this
case we obtain the following dispersion equations for the longitudinal modes
\begin{equation}
\mathcal{D}_{\parallel }(k,\omega ) =\frac{k^{2}a^{2}}{4}\delta
\varepsilon _{b\parallel }( k,\omega ) \left[ P_{1}(k,\omega ) +P_{-1}( k,\omega ) \right] ,
\label{eq:49}
\end{equation}%
where
\begin{equation}
P_{\pm 1}( k,\omega ) =\frac{\delta \varepsilon _{b\parallel
}( k,\omega \pm \omega _{0}) +\varepsilon _{\parallel }(
k,\omega ) }{\mathcal{D}_{\parallel }( k,\omega \pm \omega
_{0}) }\left[ \frac{\varepsilon _{\parallel }( k,\omega \pm
\omega _{0}) }{\varepsilon _{\parallel }( k,\omega ) }-1\right] .
\label{eq:50}
\end{equation}

The dispersion equation \eqref{eq:49} can be compared with Eqs.~\eqref{eq:41}
and \eqref{eq:42} obtained for the $\mathbf{E}_{0}\perp \mathbf{k}$
geometry. It should be emphasized that unlike the $\mathbf{E}_{0}\perp
\mathbf{k}$ geometry the dispersion equation \eqref{eq:49} (see also the more
correct relation \eqref{eq:47}) in the absence of the charged particles beam ($%
\delta \varepsilon _{b\parallel }=0$) yields the standard dispersion
equation $\varepsilon _{\parallel }(k,\omega )=0$ for the longitudinal
modes. Thus in this case the laser radiation with $\mathbf{E}_{0}\parallel
\mathbf{k}$ does not influence the dispersion properties of the plasma in
the $\mathbf{k}$ direction but it affects the dispersion relation in the
transversal direction (see Eq.~\eqref{eq:48}). This result sounds
paradoxical, considering that the plasma oscillations should be effectively
driven by the laser field in the $\mathbf{E}_{0}\parallel \mathbf{k}$
configuration. However, let us recall that the dynamics of the plasma ions
is completely neglected here. In reality, the laser radiation with $\mathbf{E%
}_{0}\parallel \mathbf{k}$\ stimulates low--frequency (typically with the ion
plasma frequency) electron--ion coupled oscillations \cite{sil73,ale84} and
the dispersion relation is not simply given by the equation $\varepsilon
_{\parallel }(k,\omega )=0$.

The simplest way to investigate the parametric TS instabilities determined by
Eq.~\eqref{eq:49} is the cold--fluid model when the dielectric functions are
given by Eq.~\eqref{eq:cold}. In this specific case the function $\Psi
_{\ell }(\zeta )$ is evaluated analytically in Appendix~\ref{sec:app1} (Eqs.~%
\eqref{eq:ap3} and \eqref{eq:ap4}) using the Newberger's summation formula
\cite{new82}. Since the function $\Psi _{\ell }(\zeta )$ decays
exponentially with $\ell $ the harmonics $E_{\parallel }(\ell )$ in Eq.~%
\eqref{eq:47} and the corresponding dispersion relations can be effectively
evaluated numerically to any order of $\ell $. Furthermore, within
cold--fluid approximation it is possible to derive a dynamical equation for
the complex amplitude of the excited waves. This is done in Appendix~\ref%
{sec:app2}, see Eq.~\eqref{eq:bp5}.

One characteristic feature of the dispersion equation \eqref{eq:49} for the
longitudinal modes in $\mathbf{E}_{0}\parallel \mathbf{k}$ geometry is the
absence of the contribution of the transverse modes. We consider first the
situation (i) (see Sec.~\ref{sec:4.1}) which corresponds to the
low--frequency ($\omega \sim \omega _{\parallel }^{\mathrm{TS}}(k)$),
long--wavelength excitations. Close to the \v{C}herenkov resonance, $\omega
\simeq ku_{b}$, we look for the solution of the dispersion equation %
\eqref{eq:49} in the form $\omega =ku_{b}+\omega _{1}$, where $\vert
\omega _{1}\vert \ll ku_{b}$. In this case the instability occurs at $%
ku_{b}\lesssim \omega _{p}$ with the growth rate ($\omega _{1}=i\gamma $)
(cf. with Eqs.~\eqref{eq:TS} and \eqref{eq:a1})
\begin{equation}
\frac{\gamma (k)}{\omega _{p}}=\left( \frac{\xi }{\gamma _{b}^{3}}\right)
^{1/2}\frac{ku_{b}}{\sqrt{\omega _{p}^{2}-(ku_{b} )^{2}}}G(k,ku_{b})
\label{eq:51}
\end{equation}%
where
\begin{equation}
G(k,\omega ) =\left\{ 1-\frac{k^{2}a^{2}}{4}\left[ P_{1}(k,\omega ) +P_{-1}(k,\omega ) \right] \right\} ^{1/2}.
\label{eq:52}
\end{equation}%
Note that Eq.~\eqref{eq:51} is valid far from the roots of the equation $\varepsilon (\omega \pm \omega _{0})=
\omega _{b}^{2}/\gamma _{b}^{3}\omega_{0}^{2}$, where $\varepsilon (\omega )=1+ \delta\varepsilon (\omega )$
is the dielectric function in a cold--fluid approximation.

The maximal growth rate is achieved at $ku_{b}\simeq \omega _{p}$. Similar to Eq.~\eqref{eq:a1}, the relation
\eqref{eq:51} is clearly invalid in this case and more rigorous treatment of the dispersion equation for the
frequency correction $\omega _{1}$\ yields a fourth order algebraic equation
\begin{equation}
\omega _{1}^{4}=\frac{\xi \omega _{p}^{3}}{2\gamma _{b}^{3}}\left[ \omega
_{1}+\frac{v_{E}^{2}}{8u_{b}^{2}}\frac{\omega _{p}\xi }{\gamma _{b}^{3}\tau
^{4}} (p_{1}+p_{-1}) \right] .
\label{eq:53}
\end{equation}%
Here
\begin{equation}
p_{\pm 1}=\frac{\varepsilon (\omega _{0}\pm \omega _{p}) }{\varepsilon (\omega _{0}\pm \omega _{p})
-\omega _{b}^{2}/\gamma_{b}^{3}\omega _{0}^{2}} ,
\label{eq:54}
\end{equation}%
and the laser frequency $\omega _{0}$ is not too close to the value $2\omega
_{p}[1+(\xi /2\gamma _{b}^{3})]$. It can be shown that the second term in
the right--hand side of Eq.~\eqref{eq:53} is systematically smaller than the
first one. Neglecting this term we arrive at the maximal growth rate $\gamma
_{\max }^{\mathrm{TS}}$ of the two--stream instability, see Eq.~%
\eqref{eq:maxTS}. Thus as expected in the situation (i) the maximal growth
rate is only weakly affected by the RF.

Next we consider high--frequency ($\omega >\omega _{\parallel }^{\mathrm{TS}%
}(k)$), short wavelength regime with $ku_{b}\gtrsim \omega _{c}$. The
nonresonant coupling similar to (iii) is determined by the intersection of
the different roots of the dispersion equations $\mathcal{D}_{\parallel
}(k,\omega \pm \omega _{0})\simeq 0$. If $\omega _{s1}$ and $\omega _{s2}$
are two different roots of the parallel dispersion function, $\mathcal{D}%
_{\parallel }(k,\omega _{s1})=\mathcal{D}_{\parallel }(k,\omega _{s2})=0$,
the above mentioned intersection of the roots simply yields $\omega
_{s1}=\omega _{s2}+2\omega _{0}$. This equation implies that both roots $%
\omega _{s1}$ and $\omega _{s2}$\ should be real which is only possible
outside the domain of the two--stream instability ($ku_{b}\gtrsim \omega _{c}$%
, see also Appendix~\ref{sec:app0}). In addition, the frequency shift $%
\Delta \omega $ determined by the relation $\omega +\omega _{0}=\omega
_{s1}+\Delta \omega $ (or alternatively $\omega -\omega _{0}=\omega
_{s2}+\Delta \omega $), is also real and can be calculated from the
dispersion equation \eqref{eq:49}. Therefore in this case the nonresonant
coupling merely shifts the real frequencies of the modes and does not cause
any instability in the beam--plasma system.

Instability may occur in situation (ii) with the resonant coupling. In this
case $\mathcal{D}_{\parallel }(k,\omega )\simeq 0$ and $\mathcal{D}%
_{\parallel }(k,\omega \pm \omega _{0})\simeq 0$. Again introducing two
different real roots $\omega _{s1}$ and $\omega _{s2}$ of the parallel
dispersion function we consider the frequency shift $\Delta \omega $\ with $%
\omega =\omega _{s2}+\Delta \omega $, such that $\vert \Delta \omega
\vert \ll \vert \omega _{s1,2}\vert $. The resonant
coupling occurs when $\omega _{s2}=\omega _{s1}\pm \omega _{0}$.
Substituting this relation and the frequency $\omega $\ into dispersion
equation \eqref{eq:49} and taking into account the smallness of the quantity
$\Delta \omega $ we obtain
\begin{equation}
\Delta \omega ^{2}\simeq \frac{k^{2}a^{2}}{4}\frac{\left[ \varepsilon (
\omega _{s2}) -\varepsilon (\omega _{s1}) \right] ^{2}}{%
\frac{\partial }{\partial \omega }\mathcal{D}_{\parallel }(k,\omega
_{s1}) \frac{\partial }{\partial \omega }\mathcal{D}_{\parallel}(k,\omega _{s2})} .
\label{eq:55}
\end{equation}%
It is seen that this expression is symmetric with respect to the exchange of the roots $\omega _{s1}$ and
$\omega _{s2}$ and yields an unstable mode if the derivatives of the longitudinal dispersion functions in
the denominator of Eq.~\eqref{eq:55} have different signs. The resonant coupling condition, $\omega _{s2}=
\omega _{s1}\pm \omega _{0}$, and the restriction on the signs of the derivatives of the longitudinal
dispersion functions in Eq.~\eqref{eq:55} reduce the possible candidates for the quantities $\omega _{s1}$
and $\omega _{s2}$. From Appendix~\ref{sec:app0} it follows that there are only three choices; a) $\omega%
_{s1}(k) =\omega_{r1}^{-}(k)$ and $\omega_{s2}(k) =\omega_{r3}(k)$ with $\omega_{r3}(k) =\omega_{0} +%
\omega_{r1}^{-}(k)$; b) $\omega_{s1}(k) =\omega_{r1}^{+}(k)$ and $\omega_{s2}(k) =\omega_{r2}(k)$ with
$\omega_{r2}(k) =\omega_{0} +\omega_{r1}^{+}(k)$; c) $\omega_{s1}(k) =\omega_{r1}^{+}(k)$ and $\omega_{s2}(k)%
=\omega_{r1}^{-}(k)$ with $\omega_{r1}^{+}(k) =\omega_{0} +\omega_{r1}^{-}(k)$. In these cases the derivatives
in the denominator of Eq.~\eqref{eq:55} have different signs and the corresponding modes are unstable. The
positions $k_{\max }$ of the maximal growth rates of these modes can be determined from the resonant coupling
condition. Using the asymptotic behavior of the roots $\omega_{r1}^{\pm}(k)$, $\omega _{r2}(k)$, and $\omega%
_{r3}(k)$ at $k\geqslant k_{c}$, Eqs.~\eqref{eq:cp1}--\eqref{eq:cp3}, respectively, we introduce four new
functions $f_{\pm}(k)$ and $h_{\pm}(k)$, which are determined through relations
\begin{eqnarray}
&&\left(
\begin{array}{c}
\omega _{r2}(k)  \\
\omega _{r3}(k)
\end{array}%
\right) =ku_{b}\mp \omega _{p}\frac{\sqrt{\xi }}{\gamma _{b}^{3/2}}h_{\pm}(k) , \label{eq:56}  \\
&&\omega _{r1}^{\pm }(k) =\pm \omega _{p}\left[ 1+\frac{\xi }{\gamma _{b}^{3}}f_{\pm}(k) \right] . \label{eq:57}
\end{eqnarray}
Then in the cases a) and b) with $\omega_{s2} =\omega_{s1} +\omega_{0}$ the resonant coupling conditions for the
determination of $k_{\max}$ in a leading order of the parameter $\xi /\gamma^{3}_{b}$ yields a pair of the
transcendental equations
\begin{equation}
ku_{b}=\omega _{0}\pm \omega _{p}\pm \frac{\sqrt{\xi }}{\gamma _{b}^{3/2}}\omega _{p}h_{\pm }(k)
\label{eq:58}
\end{equation}
which can be solved iteratively. Here the minus and plus signs are related to the cases a) and b), respectively.
Since the functions $h_{\pm}(k)$ at $k\geqslant k_{c}$ behave as $h_{\pm}(k) =1+ \mathrm{O}((\omega _{p}/ku_{b})^{2})$
(see Appendix~\ref{sec:app0}) within zero order the last term in Eq.~\eqref{eq:58} can be neglected which yields
$\kappa_{\pm} =(\omega _{0} \pm\omega _{p})/u_{b}$. Substituting this value into the arguments of the functions
$h_{\pm}(k)$ in the last term of Eq.~\eqref{eq:58} one obtains the corrections to $\kappa_{\pm}$. The maximal growth
rates are obtained from Eq.~\eqref{eq:55}, where $k=k_{\max }$. In the leading order of $\xi /\gamma _{b}^{3}$ the
result reads
\begin{equation}
\frac{\gamma _{r,\max }^{\pm }}{\omega _{p}}\simeq \frac{v_{E}}{4u_{b}}%
\left( \frac{\xi }{\gamma _{b}^{3}}\right) ^{1/4}\left\vert \frac{\tau \pm 2%
}{\tau \pm 1}\right\vert \left[ h_{\pm }(\kappa _{\pm }) \right] ^{3/2} .
\label{eq:59}
\end{equation}
For an estimate the approximate expressions $h_{\pm}(\kappa_{\pm} )\simeq 1+(1/2)(\tau \pm 1)^{-2}$ for the functions
$h_{\pm}(k)$ can be used. Here $\gamma_{r,\max }^{-}$ and $\gamma_{r,\max }^{+}$ are related to the maximal growth
rates in the regimes a) and b), respectively. Let is note that for a validity of Eq.~\eqref{eq:59} in the regime a)
the laser frequency $\omega_{0}$ should not be too close to the plasma frequency. From Eqs.~\eqref{eq:49} and \eqref{eq:55}
it is straightforward to obtain the profiles of the resonant growth rates in the regimes a) and b). Introducing the
frequency mismatch $\delta =\omega_{s2} -\omega_{s1} -\omega_{0}$ these profiles are determined by
\begin{equation}
\gamma _{r}^{\pm }(k)= \frac{1}{2} \sqrt{ 4\left(\gamma _{r,\max }^{\pm }\right) ^{2}-%
\delta^{2}(k)} .
\label{eq:new3}
\end{equation}
Here $\gamma_{r,\max }^{\pm }$ are the maximal growth rates (see Eq.~\eqref{eq:59}) which are achieved at $\delta =0$.
It is seen that the quantities $\gamma _{r}^{-}(k)$ and $\gamma _{r}^{+}(k)$ vanish at $k_{1;2} \simeq \kappa_{-} \pm
2\gamma _{r,\max }^{-}/u_{b}$ and $k_{3;4} \simeq \kappa_{+} \pm 2\gamma _{r,\max }^{+}/u_{b}$, respectively.

Consider now the regime c) when $\omega_{s2} =\omega_{s1} -\omega_{0}$. In this case the resonant coupling condition reads
\begin{equation}
\omega _{0}-2\omega _{p}=\omega _{p}\frac{\xi }{\gamma _{b}^{3}}\left[f_{+}(k) +f_{-}(k) \right] .
\label{eq:new1}
\end{equation}
It is clear that this relation can be satisfied only at $\omega_{0} >2\omega_{p}$ since the functions $f_{\pm}(k)$
are positive (see Appendix~\ref{sec:app0}). On the other hand $\omega_{0}$ should be sufficiently close to
$2\omega_{p}$ because the dimensionless parameter $\xi /\gamma_{b}^{3}$ is small. Assuming for instance  $\tau%
-2\ll \xi /\gamma _{b}^{3}$ and employing the results of the Appendix~\ref{sec:app0} the solution of Eq.~\eqref{eq:new1}
reads as $k_{0}u_{b}/\omega _{p}\simeq (\xi /\gamma _{b}^{3})^{1/2}(\tau -2)^{-1/2}\gg 1$. This value of $k$ determines
the position of the maximal growth rate in c) which is obtained from Eq.~\eqref{eq:55} and is given by
\begin{equation}
\frac{\gamma _{r,\max }}{\omega _{p}}\simeq \frac{v_{E}}{u_{b}}\frac{\tau -2%
}{\tau }\ll \frac{v_{E}}{u_{b}}\frac{\xi }{\tau \gamma _{b}^{3}} .
\label{eq:new2}
\end{equation}
Confronting this relation with the growth rates $\gamma _{r,\max }^{\pm }$ we conclude that $\gamma_{r,\max } \ll%
\gamma _{r,\max }^{\pm }$. In addition it should be noted that in general $k_{0} \gg \kappa_{\pm}$ and the growth
rate $\gamma_{r,\max}$ may be strongly shifted towards very large $k$ values.

Unlike $\mathbf{E}_{0}\perp \mathbf{k}$ geometry the beam--plasma and the laser--plasma parametric modes are strongly
coupled here. As a result the resonant modes~\eqref{eq:59} and \eqref{eq:new2} depend essentially on the laser
intensity and the beam density.

Let us now compare the growth rates in the regimes a) and b) for the resonant unstable modes with the quantity
$\gamma_{\max }^{\mathrm{TS}}$ assuming, for simplicity $\tau \gg 1$ ($\omega _{0}\gg \omega _{p}$). In this case
$\gamma^{+}_{r,\max} \simeq \gamma^{-}_{r,\max}$. It is seen that $\gamma^{\pm}_{r,\max}$ exceed the growth rate
$\gamma _{\max }^{\mathrm{TS}}$ of the two--stream instability at sufficiently intense RF, $v_{E}/c>3.02(\gamma%
_{\max }^{\mathrm{TS}}/\omega _{p})^{1/4}$. It is clear that this condition requires very low density beams
(compared to $n_{e}$) and is increasingly difficult to fulfill with increasing $n_{b}$. Also the growth rates in
Eq.~\eqref{eq:59} should be compared with the growth rates of the resonant longitudinal (Eq.~\eqref{eq:a8}) and
the transversal (Eqs.~\eqref{eq:x3} and \eqref{eq:x5}) modes as well as with the nonresonant longitudinal modes
(Eqs.~\eqref{eq:a9} and \eqref{eq:a10}) excited in $\mathbf{E}_{0}\perp \mathbf{k}$ geometry. Again assuming, for
simplicity the $\omega _{0}\gg \omega _{p}$ limit we conclude that the unstable modes grow much faster in this geometry,
where the parametric effects are more pronounced. However, it should be emphasized that the resonant unstable modes
in $\mathbf{E}_{0}\perp \mathbf{k}$ geometry are only effectively excited starting with the threshold frequency
$\omega _{0}\simeq 2\omega _{p}$ of the laser radiation. Below this threshold (with $\omega_{0}\lesssim 2\omega_{p}$)
only the unstable mode \eqref{eq:x7} in the long wavelength domain and the modes \eqref{eq:a9}, \eqref{eq:a10},
and \eqref{eq:59} in the short wavelength domain are excited in $\mathbf{E}_{0}\perp \mathbf{k}$ and $\mathbf{E}_{0}%
\parallel \mathbf{k}$ configurations, respectively.

\section{Numerical treatment}
\label{sec:6}

Using the theoretical findings of Secs.~\ref{sec:3}--\ref{sec:5}, we present here the results of
our numerical calculations of the growth rates for the longitudinal and the transversal unstable modes
assuming the transverse ($\mathbf{E}_{0}\perp \mathbf{k}$) and the parallel ($\mathbf{E}_{0}\parallel
\mathbf{k}$) configuration of the RF amplitude $\mathbf{E}_{0}$ with respect to the wave vector $\mathbf{k}$.
The calculations have been done for an electron beam with a small dimensionless density parameters $\xi =(\omega
_{b}/\omega _{p})^{2}=n_{b}/n_{e}=0.1$ and $\xi =0.3$ and for a relativistic factor $\gamma _{b}=5$.
For the laser intensity parameter $\alpha =v_{E}^{2}/c^{2}$ we have adopted the values $\alpha =0,\, 0.01,\,
0.1,$ and $0.2$. It is convenient to represent the laser intensity parameter in the form $\alpha =(I_{L}/%
I_{0})\lambda _{0}^{2}$, where $I_{0}=1.37\times 10^{18}$ W\,$\mu$m$^{2}$/cm$^{2}$ and the wavelength
($\lambda _{0}$) and the intensity ($I_{L}$) of the laser field are measured in units $\mu$m and W/cm$^{2}$,
respectively. The laser frequency is measured in the units of the plasma frequency, $\tau =\omega _{0}/\omega _{p}>1$.
In our numerical calculations this parameter varies in a wide interval,
$1.2\leqslant \tau \leqslant 4$. Throughout in this section the growth rates are measured in units of plasma
frequency $\omega_{p}$ and are calculated as a function of $ku_{b}/\omega_{p}$ for several laser intensities
and frequencies. Note that the chosen parameters both for electron beam and RF are typical for FIS for inertial
confinement fusion \cite{deu96}. Assuming, for instance, radiation field with $\lambda_{0} =0.5$~$\mu$m, the
parameter $\alpha =0.2$ corresponds to the intensity $I_{L}\simeq 10^{18}$~W/cm$^{2}$ of the RF.

First we consider the transverse geometry with $\mathbf{E}_{0}\perp \mathbf{k}$. In this case the basic properties
of the dispersion relations for the beam--plasma system have been studied in Sec.~\ref{sec:4}. In general we
have found that the simplified treatments of Secs.~\ref{sec:4} and \ref{sec:5} agree qualitatively well with
the exact numerical solutions. However, it is clear that these simplified treatments are not capable to resolve
all details and the branches of the spectrum of the unstable modes in the $\omega$--$k$ plane.

\begin{figure*}[tbp]
\includegraphics[width=80mm]{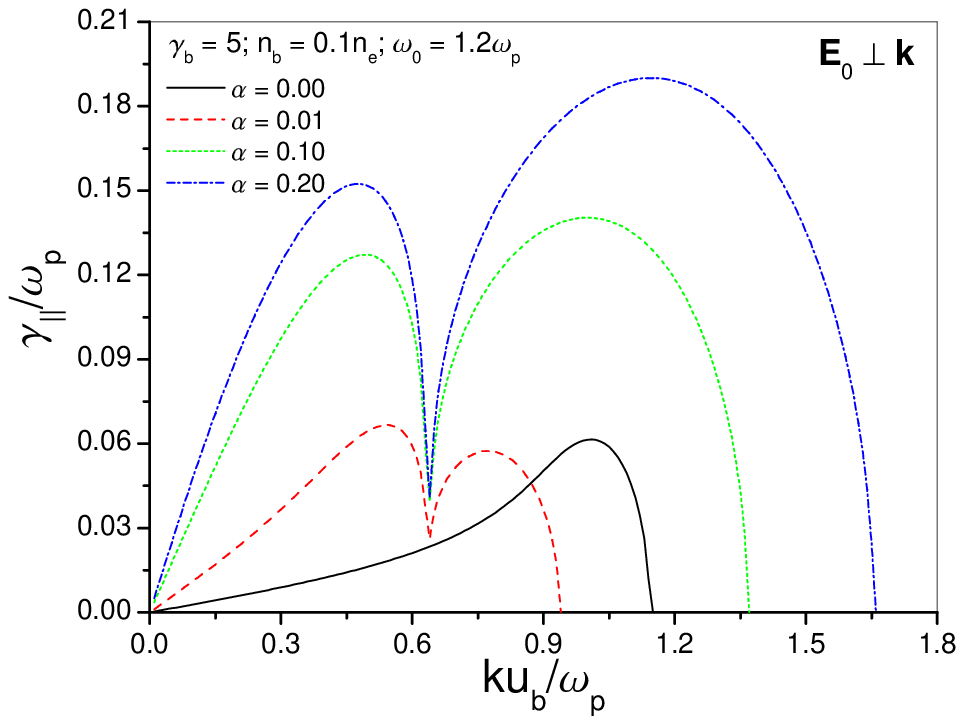}
\includegraphics[width=80mm]{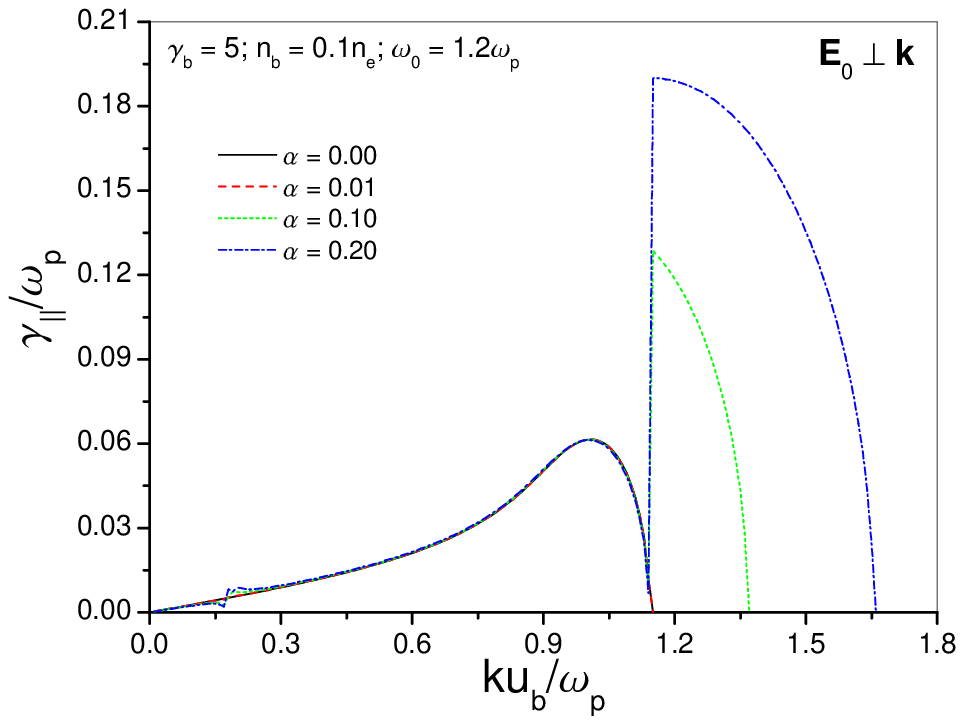}
\caption{(Color online) The growth rate $\gamma _{\parallel }$ (in units of plasma frequency $\omega _{p}$)
of the longitudinal unstable modes in terms of $ku_{b}/\omega _{p}$ obtained by numerical solution of the
dispersion equation~\eqref{eq:41} in $\mathbf{E}_{0}\perp \mathbf{k}$ configuration for $\gamma _{b}=5$,
$n_{b}=0.1n_{e}$, $\omega _{0}=1.2\omega _{p}$, $\alpha =0$ (solid line), $\alpha =0.01$ (dashed line),
$\alpha =0.1$ (dotted line), $\alpha =0.2$ (dash--dotted line). The left and the right panels correspond
to the branches~I, II and III, IV, respectively, introduced in the text.}
\label{fig:1}
\end{figure*}

Within $\mathbf{E}_{0}\perp \mathbf{k}$ geometry we now consider the case of the longitudinal unstable modes
($\gamma =\gamma_{\parallel}$) when the dispersion relations are determined by Eq.~\eqref{eq:41}. As mentioned
in Sec.~\ref{sec:4.1} this is a regime when the purely two--stream and the parametric modes are only weakly
coupled and their spectra are well separated in the $\omega$--$k$ plane. Therefore the different modes (beam--plasma
or parametric) are only weakly sensitive either to the electron beam parameters or the intensity of the RF. From Eq.~\eqref{eq:41}
it is seen (see also the simplified version of this relation, Eq.~\eqref{eq:a5}) that in a cold--fluid
approximation there are ten solutions of this equation, but only some of them correspond to the unstable
modes with $\gamma >0$. In addition, these unstable modes are excited with different real frequencies. To
demonstrate this feature in Fig.~\ref{fig:1} the growth rates are shown for the laser frequency $\omega_{0} =
1.2\omega_{p}$. In this case we have found numerically that there are only three solutions which correspond
to the unstable modes, and as an example two solutions are shown in the left and right panels of Fig.~\ref{fig:1}.
The different curves correspond to the laser dimensionless intensities $\alpha =0.01$ (dashed lines), $\alpha =0.1$
(dotted lines), and $\alpha =0.2$ (dash--dotted lines). The solid lines with $\alpha =0$ represent the growth
rate of the standard two--stream instability, Eq.~\eqref{eq:TS}. Two panels of Fig.~\ref{fig:1} correspond to
the modes with different real frequencies $\omega_{r}$. We denote tentatively the solutions with different
$\omega_{r}$ as the branches~I, II, III etc. In the left panel of Fig.~\ref{fig:1} up to the value $k_{2}c%
/\omega_{p} =(\tau^{2}-1)^{1/2} \simeq 0.6$ ($k=k_{2}$ corresponds to the vanishing frequency mismatch, $\delta =0$,
introduced in Sec.~\ref{sec:4.1}) the real frequency is $\omega_{r} =0$ (branch~I), while at $k_{2}\leqslant %
k\leqslant k_{1}$ it is given by $\omega_{r} = \omega_{g}(k)$ (branch~II). In Fig.~\ref{fig:1} the growth rates
sharply tend to zero at $k= k_{1}$. [For the approximate definition of the quantity $k_{1}$ see paragraph above
Eq.~\eqref{eq:a10}]. The spectrum $\omega_{g}(k)$ corresponds to the real frequency of the nonresonant longitudinal
modes derived approximately in Sec.~\ref{sec:4.1} for the negative frequency mismatch ($\delta <0$). In the
approximate form it is given by Eq.~\eqref{eq:a10}, where, however, the minus sign has to be replaced by the
plus sign. In the approximate treatment of Sec.~\ref{sec:4.1}, Eqs.~\eqref{eq:a9} and \eqref{eq:a10} correspond
to the branches I and II, respectively. Thus the boundary between the branches~I and II is determined by $\delta =0$
(or $k=k_{2}$). In the branch~I the growth rate at $k\lesssim k_{2}$ increases almost linearly with $k$, $\gamma (k)%
\simeq kv_{E}[2(\tau^{2} -1)]^{-1/2}$ in agreement with Eq.~\eqref{eq:a9}. Finally, in Fig.~\ref{fig:1} (right panel)
the real frequencies of the modes at $0\leqslant k\leqslant k_{c}$ and $k_{c}\leqslant k\leqslant \max[k_{c};k_{1}]$
coincide with the real frequency of the standard two--stream instability, $\omega_{r}= \omega^{\mathrm{TS}}_{\parallel}(k)$
(branch~III, see Eq.~\eqref{eq:TS}) and $\omega_{r}= \omega_{0} +\omega_{\perp}(k)$ (branch~IV), respectively. Here
$\omega_{\perp}(k)$ is given by Eq.~\eqref{eq:tr}. Let us note that the boundary $k=k_{2}$ between~I and II does not
depend on the intensity of the RF while the upper boundary $k=k_{1}$ of the branches~II and IV (at $k_{1} >k_{c}$)
is shifted towards shorter wavelengths roughly as $k_{1}\simeq k_{2}[1+ \mathrm{O}(\alpha^{1/3})]$ with increasing laser
intensity. It is clear that at smaller intensity of the RF when $k_{1} <k_{c}$ the branch~IV disappears. As a general
rule we observe that at $k_{c}\leqslant k\leqslant k_{1}$ the growth rates in the branch~IV are the part of the branch~II,
except the case of the low intensity RF with $k_{1} <k_{c}$ (cf. the dashed curves in Fig.~\ref{fig:1} with $\alpha =0.01$).
In this low--intensity limit the branch~IV disappears while the branch~III is nearly the same as the standard branch for
the two--stream instability (solid curves in Fig.~\ref{fig:1}). Thus, at $\omega_{0}\gtrsim \omega_{p}$ and at small
intensities of the RF ($k_{1} <k_{c}$) the parametric two--stream instability occurs in the branch~III with the growth
rate $\simeq \gamma^{\mathrm{TS}} (k)$ which is only weakly affected by the RF. At higher intensities of the RF (with
$k_{1} >k_{c}$) a new unstable branch~IV is formed. The branches~I, II and IV are formed due to the parametric excitations
and are almost insensitive to the electron beam.

\begin{figure*}[tbp]
\includegraphics[width=80mm]{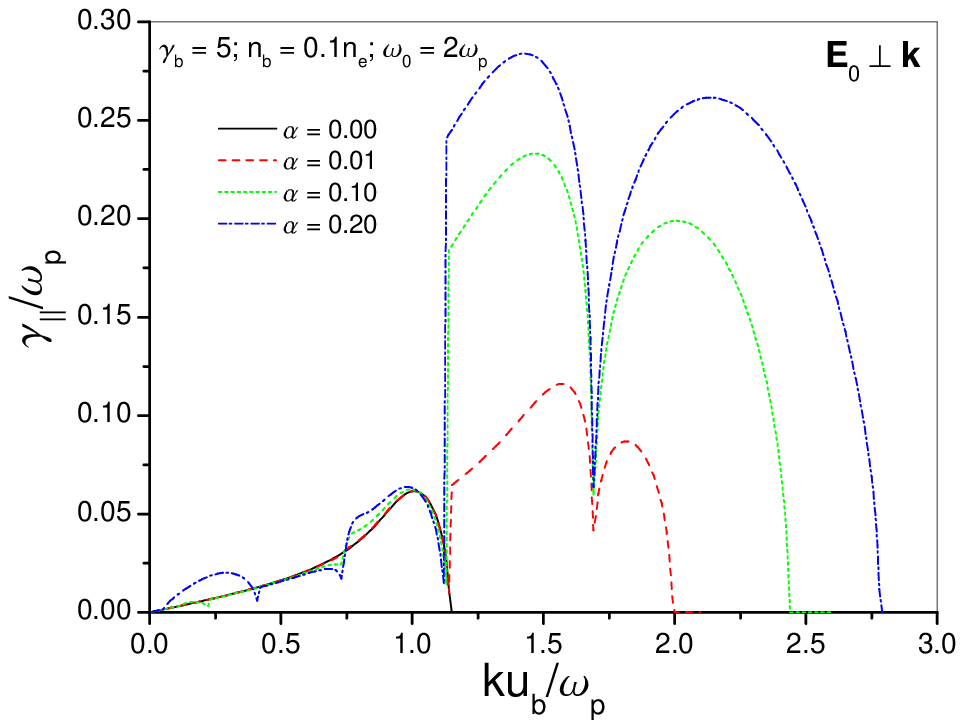}
\includegraphics[width=79mm]{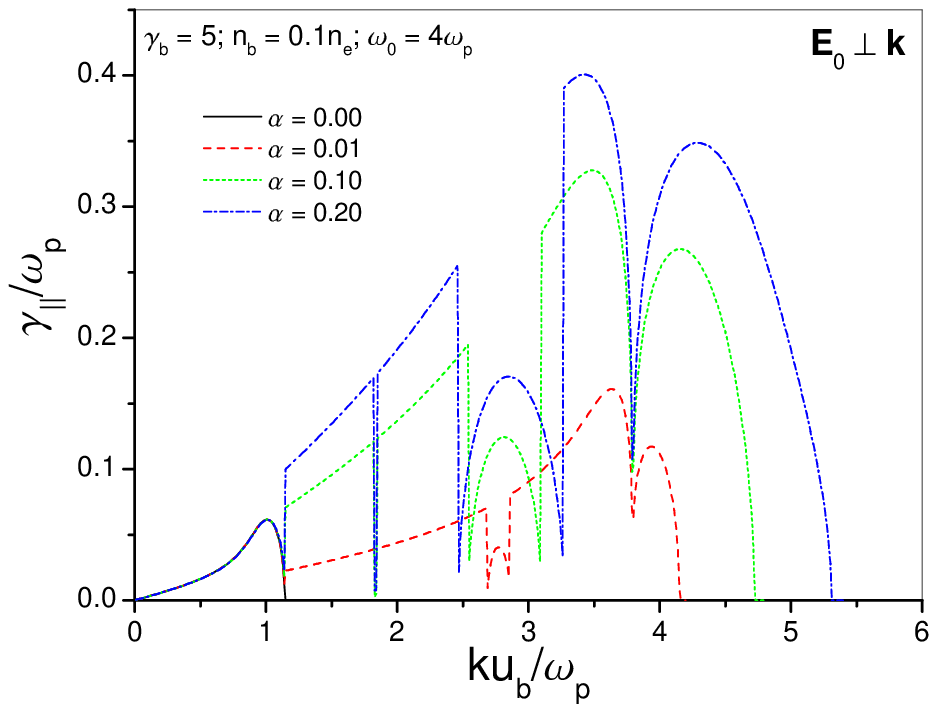}
\caption{(Color online) Same as in Fig.~\ref{fig:1} but for the higher laser frequencies $\omega_{0} =2\omega_{p}$
(left panel) and $\omega_{0} =4\omega_{p}$ (right panel). Note the different scales in the left and the right
panels.}
\label{fig:2}
\end{figure*}

\begin{figure*}[tbp]
\includegraphics[width=80mm]{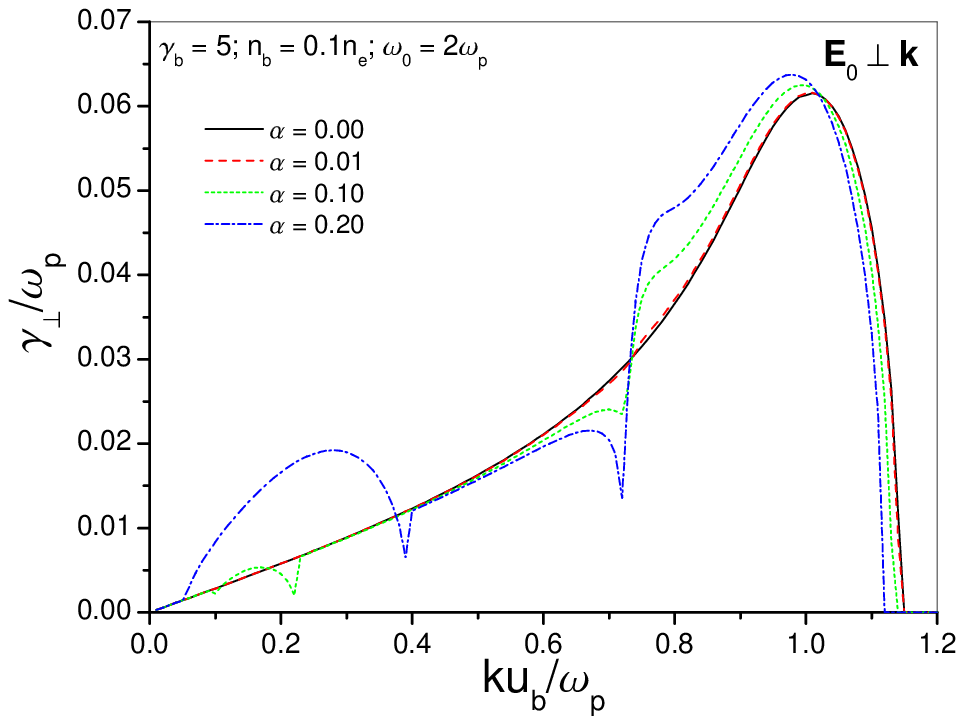}
\includegraphics[width=80mm]{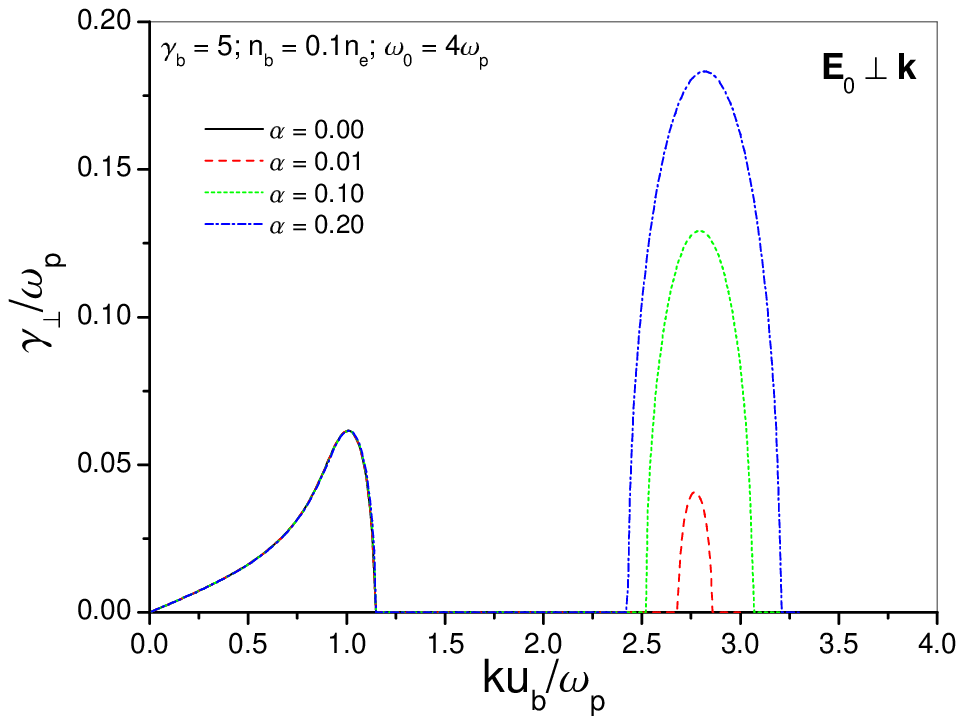}
\caption{(Color online) The growth rate $\gamma _{\perp}$ (in units of plasma frequency $\omega _{p}$) of the transversal
unstable modes in terms of $ku_{b}/\omega _{p}$ obtained by numerical solution of the dispersion equation~\eqref{eq:42}
in $\mathbf{E}_{0}\perp \mathbf{k}$ configuration for $\gamma _{b}=5$, $n_{b}=0.1n_{e}$, $\omega _{0}=2\omega _{p}$
(left panel), $\omega _{0} =4\omega _{p}$ (right panel), and for $\alpha =0$ (solid line), $\alpha =0.01$ (dashed line),
$\alpha =0.1$ (dotted line), $\alpha =0.2$ (dash--dotted line). Note the different scales in the left and the right
panels.}
\label{fig:3}
\end{figure*}

Next in Fig.~\ref{fig:2} the longitudinal growth rates are shown for the higher laser frequencies $\omega_{0} =2\omega_{p}$
(left panel) and $\omega_{0} =4\omega_{p}$ (right panel). As expected the different branches shown in Fig.~\ref{fig:1}
are mixed with increasing $\omega_{0}$ and, in addition, more and more new branches for the unstable modes are formed.
As an example in Fig.~\ref{fig:2} only two solutions of the dispersion equation~\eqref{eq:41} are shown which involve
the basic features of the branches~I, II, III, and IV introduced above. Note that in these particular examples with
higher laser frequencies, $k_{2}\simeq \omega_{0}/c$ exceeds the upper boundary $k_{c}$ of the two-stream instability,
and therefore $k_{1} >k_{c}$ for an arbitrary intensity of the RF. As has been pointed out above the approximate growth
rates given by Eqs.~\eqref{eq:a9} and \eqref{eq:a10} are not capable in these regimes to resolve all specific branches
shown in Fig.~\ref{fig:2}. The most important feature shown in Fig.~\ref{fig:2} is that the domain of the instability
in the $k$--space is broadened accompanied by an increase of the maximal growth rate with increasing $\omega_{0}$. It
is also noteworthy the formation of the resonant growth rate in the spectrum of the unstable modes at $\omega_{0}\gtrsim
2\omega_{p}$ derived approximately in Sec.~\ref{sec:4.1}, see Eqs.~\eqref{eq:a6}--\eqref{eq:a8}. Let us recall that
the resonant unstable mode is not excited at $\omega_{0}< 2\omega_{p}$ (see Sec.~\ref{sec:4.1}) and hence this mode
is not visible on Fig.~\ref{fig:1}. As predicted by Eqs.~\eqref{eq:a6}--\eqref{eq:a8} the resonant coupling at $\omega_{0}=
2\omega_{p}$ (Fig.~\ref{fig:2}, left panel) is only weakly pronounced with the maximal growth rate $\gamma_{r,\max}
/\omega_{p} \simeq \alpha /6$ and $k_{r,\max} \simeq (\omega_{p}/c)(\alpha /3)^{1/2}$ while at higher frequency
$\omega_{0} =4\omega_{p}$ (Fig.~\ref{fig:2}, right panel) it is strongly increased and is shifted towards the short
wavelengths, $\gamma_{r,\max}/\omega_{p} \sim (\alpha\tau/4)^{1/2}$ with $k_{r,\max}\sim \omega_{0}/c$. In Fig.~\ref{fig:2}
(right panel) the resonant growth rate is determined by the curve with the maximum located around $ku_{b}/\omega_{p}
\sim 3$.

\begin{figure*}[tbp]
\includegraphics[width=80mm]{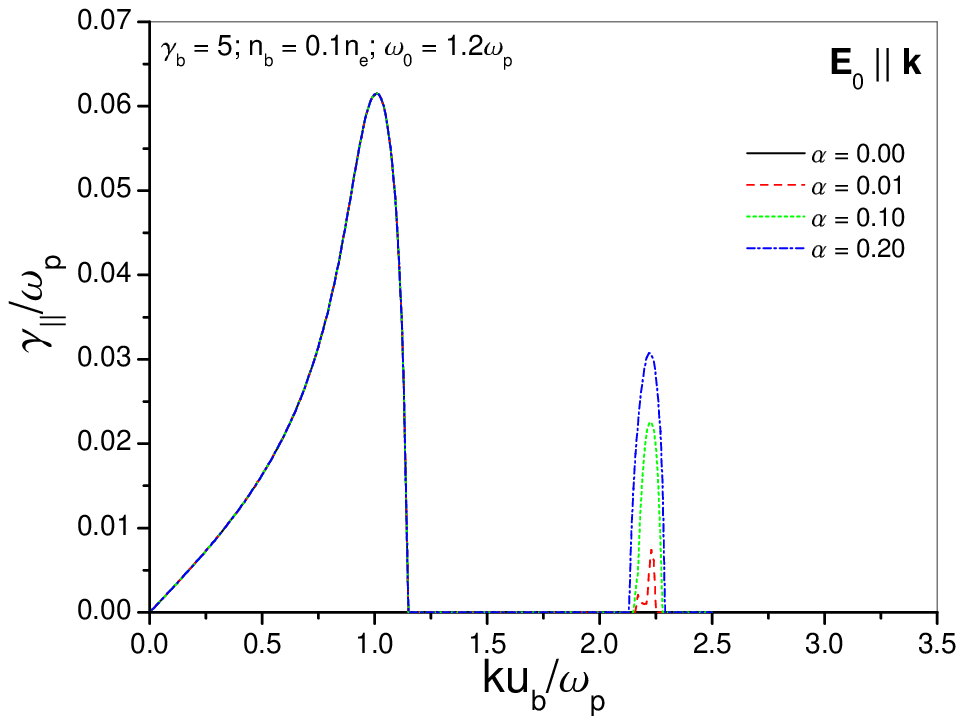}
\includegraphics[width=80mm]{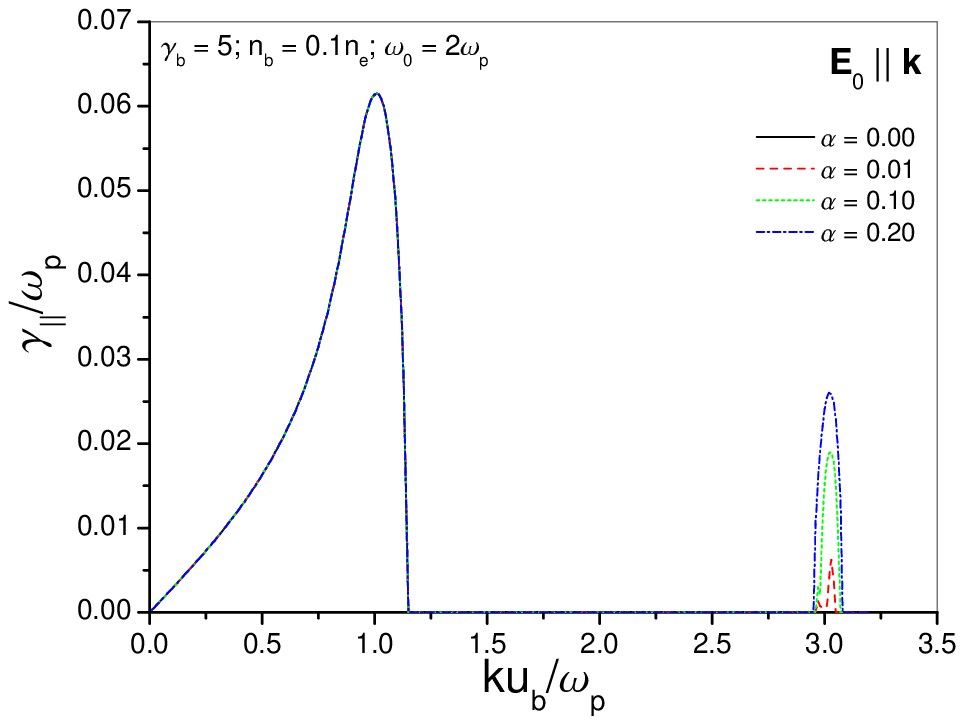}
\caption{(Color online) The growth rate $\gamma _{\parallel}$ (in units of plasma frequency $\omega _{p}$) of the longitudinal
unstable modes in terms of $ku_{b}/\omega _{p}$ obtained by numerical solution of the dispersion equation~\eqref{eq:49}
with Eq.~\eqref{eq:50} in $\mathbf{E}_{0}\parallel \mathbf{k}$ configuration for $\gamma _{b}=5$, $n_{b}=0.1n_{e}$, $\omega%
_{0}=1.2\omega _{p}$ (left panel), $\omega _{0} =2\omega _{p}$ (right panel), and for $\alpha =0$ (solid line), $\alpha =0.01$
(dashed line), $\alpha =0.1$ (dotted line), $\alpha =0.2$ (dash--dotted line).}
\label{fig:4}
\end{figure*}

\begin{figure*}[tbp]
\includegraphics[width=80mm]{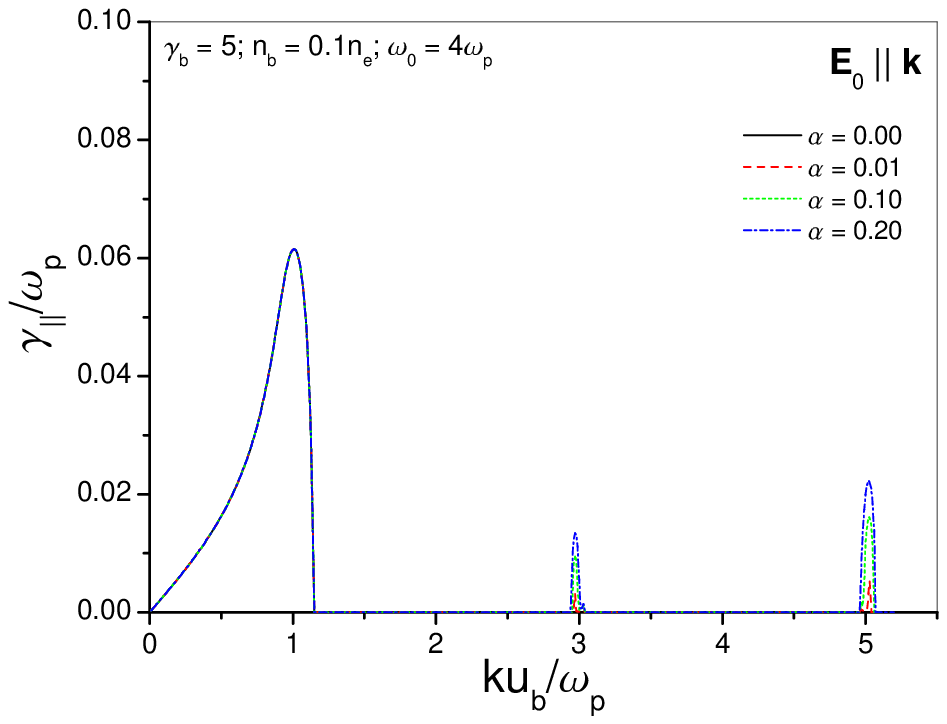}
\includegraphics[width=80mm]{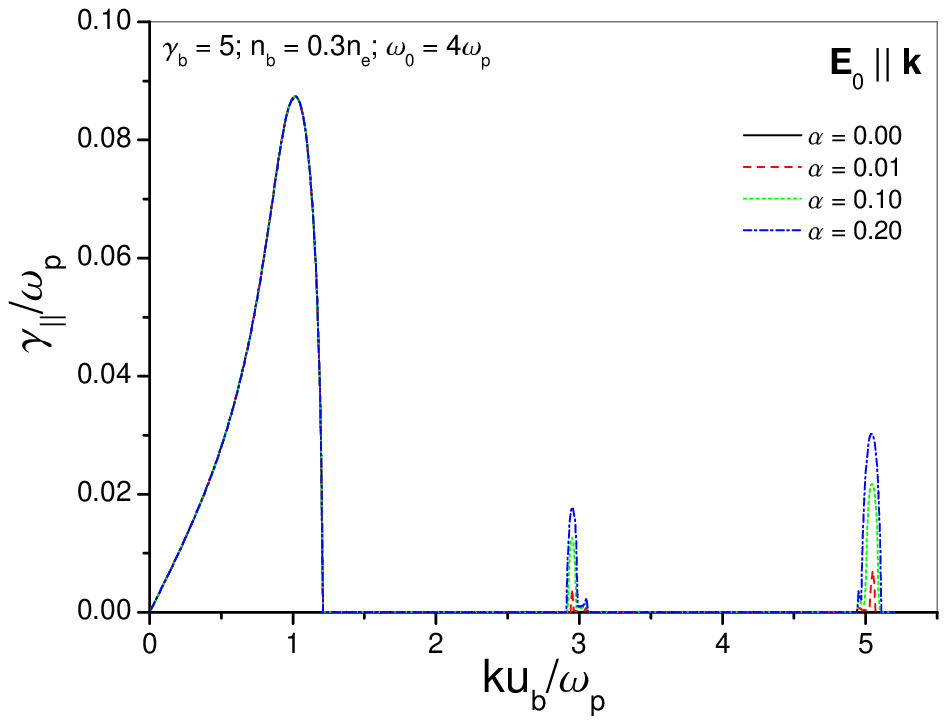}
\caption{(Color online) Same as in Fig.~\ref{fig:4} but for $\omega _{0} =4\omega _{p}$, $n_{b}=0.1n_{e}$ (left panel) and
$n_{b}=0.3n_{e}$ (right panel).}
\label{fig:5}
\end{figure*}

The growth rates for the transversal unstable modes are demonstrated in Fig.~\ref{fig:3} for the $\mathbf{E}_{0}\perp%
\mathbf{k}$ geometry and for $\omega_{0} =2\omega_{p}$ (left panel) and $\omega_{0} =4\omega_{p}$ (right panel). These
growth rates are obtained by the numerical solution of the dispersion equation~\eqref{eq:42} for the transversal modes.
The results for the smaller laser frequencies (with $\omega_{0}< 2\omega_{p}$) are not shown in Fig.~\ref{fig:3}. This
is because only the nonresonant modes with the growth rates approximately given by Eqs.~\eqref{eq:x6} and \eqref{eq:x7}
are possible in this case as discussed in Sec.~\ref{sec:4.2}. We have found numerically that in this frequency regime
the growth rate only weakly deviates from the growth rate $\gamma^{\mathrm{TS}} (k)$ (see Eq.~\eqref{eq:TS}) of the
standard two--stream instability which is supported by the approximate Eq.~\eqref{eq:x7}. Thus in Fig.~\ref{fig:3}
the nearly resonant and the resonant cases are shown with $\omega_{0} =2\omega_{p}$ and $\omega_{0} =4\omega_{p}$,
respectively. In the first case the high--frequency ($\omega\simeq \omega_{0} -\omega_{p}$, see Sec.~\ref{sec:4.2})
resonant mode is not yet formed but it may interfere with the two--stream mode essentially changing the growth rate,
see Fig.~\ref{fig:3} (left panel). And this effect is increasing with the laser intensity. In the second case the resonant
mode is well separated from the two--stream mode and forms (see Fig.~\ref{fig:3}, right panel) an isolated maximum at
$k_{\max}$ determined approximately by Eq.~\eqref{eq:x2} (or roughly $k_{\max}\simeq (\omega_{p}/c) [\tau (\tau -1)]^{1/2}$).
The maximal growth rate of the resonant mode is well described by Eq.~\eqref{eq:x3}. It is seen that the position of
the maximum is almost independent from the laser intensity while the maximum growth rate increases as $\sim \alpha^{1/2}$
with the parameter $\alpha$. It is also noteworthy the dependence of the growth rate $\gamma_{\perp} (k)$ of the resonant
mode on the frequency $\omega_{0}$ of the laser field for fixed plasma density and RF intensity. The maximal growth rate
$\gamma_{r,\max}$ increases with $\omega_{0}$ and is shifted towards larger $k$ achieving the maximal value at $\omega_{0}%
\simeq 3.4\omega_{p}$. For larger frequencies the maximal growth rate is scaled as $\gamma_{r,\max} \sim [I_{L}(\omega_{p}%
/\omega_{0})]^{1/2}$ and falls with $\omega_{0}$.

Finally in Figs.~\ref{fig:4} and \ref{fig:5} the growth rate $\gamma _{\parallel}$ of the longitudinal unstable modes
excited in the case $\mathbf{E}_{0}\parallel \mathbf{k}$ are shown. These results have been obtained by numerical
solution of the dispersion equation~\eqref{eq:49} with Eq.~\eqref{eq:50} for $\omega_{0}=1.2\omega _{p}$, $\omega_{0}
=2\omega _{p}$ (Fig.~\ref{fig:4}) and $\omega_{0}=4\omega _{p}$ (Fig.~\ref{fig:5}). As mentioned in Sec.~\ref{sec:5}
the coupling between the parametric and the two--stream modes may be very effective in this configuration which is supported
by the analytical results obtained in Sec.~\ref{sec:5}. The growth rate~\eqref{eq:new2} of the resonant unstable mode
in the regime c) is much smaller than $\gamma^{\pm}_{r,\max}$ and is not shown here. In both panels of Fig.~\ref{fig:4}
the frequency of the RF is rather small and hence only the right--side resonant mode with $\gamma^{+}_{r,\max}$ is
excited at $\kappa_{+}u_{b}\simeq \omega_{0}+\omega_{p}$ (see Eq.~\eqref{eq:59}). The left--side resonant mode with
$\gamma^{-}_{r,\max}$ is formed at $\kappa_{-}u_{b}\simeq \omega_{0}-\omega_{p}$ and in Fig.~\ref{fig:4} it is merged
with the two--stream mode and is not distinguishable. To gain more insight in Fig.~\ref{fig:5} we demonstrate the
growth rates for the larger laser frequency $\omega_{0}=4\omega _{p}$ assuming that $n_{b} =0.1n_{e}$ (left panel) and
$n_{b} =0.3n_{e}$ (right panel). Now with increasing laser frequency $\omega_{0}$ the left--side resonant mode is
clearly visible in Fig.~\ref{fig:5} and the corresponding growth rate $\gamma^{-}_{r,\max}$ is smaller than
$\gamma^{+}_{r,\max}$ as predicted by Eq.~\eqref{eq:59}. The domains where the growth rates of the left--side and
the right--side resonant modes are nonzero can be approximated as $\Delta k_{\pm }\simeq 4\gamma _{r,\max }^{\pm }/u_{b}$,
see Eq.~\eqref{eq:new3}. Furthermore, both $\gamma^{\pm}_{r,\max}$ and $\Delta k_{\pm }$ increase with electron density
as shown in Fig.~\ref{fig:5} (right panel). Note that the growth rates $\gamma^{\mathrm{TS}}_{\max}$ and $\gamma^{\pm}_{r,\max}$
increase approximately as $\sim n_{b}^{1/3}$ and $\sim n_{b}^{1/4}$ with the beam density $n_{b}$, respectively.
In the parameter regimes shown in Figs.~\ref{fig:4} and \ref{fig:5} the electron beam is somewhat dense and the
condition $\gamma^{\pm}_{r,\max} >\gamma^{\mathrm{TS}}_{\max}$ requires relativistic intensities for the RF (with
$v_{E} >c$) which cannot be fulfilled in the present approximation.

In the numerical examples shown in Figs.~\ref{fig:1}--\ref{fig:5} the parametric instabilities are comparable or even
stronger than the standard two--stream instability. The stronger effect is expected rather in the case of the transversal
polarization $\mathbf{E}_{0}\perp \mathbf{k}$ of the laser field both for the longitudinal and transversal modes.
In addition, the $k$--domain of the parametric instability is comparable or even larger than the range $k_{c}$ of the
standard two--stream instability. Although an effective coupling between laser--plasma and beam--plasma modes is
expected in the case of the parallel polarization of the laser field, $\mathbf{E}_{0}\parallel \mathbf{k}$, the resulting
growth rates and their $k$--domains are in general essentially smaller than those obtained for the transversal polarization.
This is not surprising because the effects of the laser radiation and the REB on the dispersion properties of plasma are
treated here perturbatively when the coupling between both effects is rather weak.

\section{Summary and concluding remarks}
\label{sec:7}

In this paper, we have presented a theoretical study of the growth rates of the unstable modes excited simultaneously
by a laser field and a relativistic beam of charged particles moving in an isotropic plasma. The laser field
is treated in the long wavelength limit (dipole approximation) and the plasma particles are considered nonrelativistic.
In addition, an ultrarelativistic beam of the charged particles is considered and the influence of the laser field
on the beam is neglected. The dynamics of the beam--plasma system in the presence of the RF is studied by the linearized
relativistic and nonrelativistic Vlasov kinetic equations for the distribution functions of the beam and the plasma,
respectively, as well as by the linearized Maxwell equations for the electromagnetic fields. The full electromagnetic
response of the system is derived in terms of the conductivity tensor of the system involving all harmonics of the RF.
After a general introduction to the theoretical model in Sec.~\ref{sec:2}, the dispersion relations of the modes
are considered in Secs.~\ref{sec:3} and \ref{sec:4}. It is shown that in general the longitudinal and transversal modes
are parametrically coupled due to the presence of the RF. As a result the dispersion equation of the modes represents
a secular equation for each harmonic of the electromagnetic field. Assuming, however, nonrelativistic laser intensities
in Secs.~\ref{sec:4} and \ref{sec:5} we have considered only the lowest (zero, first and second) harmonics of the fields
truncating the secular dispersion equation on the second order of the amplitude of the RF. The dispersion equations
derived in these sections led to a detailed presentation, in Sec.~\ref{sec:6}, of a collection of data through figures
on the growth rates. For numerical calculations we have chosen $\gamma_{b} =5$ which pertains to the FIS relativistic
electron beam in the typical 1--2~MeV energy range of practical interest.

Explicit calculations have been done within cold--fluid approximation both for the beam and the plasma neglecting the
low--frequency dynamics of the plasma ions. Furthermore, the beam drift velocity is parallel to the wave vector $\mathbf{k}$
of the excitations. Two particular cases of the transverse ($\mathbf{E}_{0} \perp \mathbf{k}$) and the longitudinal
($\mathbf{E}_{0} \parallel \mathbf{k}$) polarizations of the RF have been studied in detail in Secs.~\ref{sec:4} and
\ref{sec:5}, respectively. For the longitudinal and the transversal unstable modes we have identified some domains in the
$\omega$--$k$ plane corresponding to the resonant and the nonresonant instabilities occurring due to the parametric coupling
of the different modes. Analytical expressions for the relevant growth rates have been derived which are well supported
by our numerical calculations in Sec.~\ref{sec:6}. These analytical results go beyond those obtained previously in
Refs.~\cite{ali66,gor66,ali79} (see also Refs.~\cite{sil73,ale84}) and in Refs.~\cite{bre04,bre05,bre06} for the purely
parametric and the beam--plasma unstable modes, respectively. In the course of this study, we have also derived in
Appendix~\ref{sec:app2} a dynamical equation for the complex amplitude of the excited modes in $\mathbf{E}_{0}\parallel
\mathbf{k}$ configuration without any restriction on the number of harmonics. When the initial conditions are specified
this equation may be useful in analyzing the beam--plasma parametric instabilities beyond a weak RF limit considered here.

It was shown that in the case of the transverse $\mathbf{E}_{0} \perp \mathbf{k}$ polarization of the RF the longitudinal
and the transversal modes are coupled due to the RF and can be unstable. The purely beam--plasma and the parametric unstable
modes are only weakly coupled in this case. With increasing the laser frequency $\omega_{0}$ the new branches of the
parametrically unstable modes are excited. Furthermore, as demonstrated in Figs.~\ref{fig:1}--\ref{fig:3} the growth rates
of these modes (as well as the corresponding domain in the $k$--space) essentially increase with $\omega_{0}$ and the laser
intensity and exceed the growth rate $\gamma^{\mathrm{TS}}$ of the standard two--stream instability. In the case of the
parallel polarization (with $\mathbf{E}_{0} \parallel \mathbf{k}$) the longitudinal and the transversal modes are decoupled
and the instability occurs mainly for the former modes. The purely beam--plasma and the parametric unstable modes are now
strongly coupled and the whole instability domain in the $k$--space is split (at $\omega_{0} >2\omega_{p}$) into three major
subdomains. In the long wavelength subdomain with $0<k<k_{c}$ the instability is similar to the two--stream one while at
$ku_{b} \simeq \omega_{0}\pm \omega_{p}$ the instability essentially depends on both the beam and the RF parameters (see
Figs.~\ref{fig:4} and \ref{fig:5}). Finally, it was shown that the growth rates are larger in the case of the transversal
polarization $\mathbf{E}_{0}\perp \mathbf{k}$ of the laser field.

Going beyond the present approach, which is based on several approximations and assumptions, we can envisage a number of
improvements. These include (i) the effects of the low--frequency dynamics of the plasma ions which are completely neglected
here provided that the obtained frequencies are much higher than the ionic frequencies; (ii) thermal effects both for the
beam and the plasma. In principle in the case of a standard beam--plasma instability the growth rate can be reduced by
these effects \cite{mik74}. However, in the case of a relativistic beam the thermal momentum spreads of the beam particles
in the directions parallel or transverse to the beam velocity has only little influence on the instability \cite{bret06}.
Therefore, these effects are important in the case of the nonrelativistic beams as, for instance, in the experiments with
heavy ion beams interacting with a laser irradiated plasma \cite{sto96,rot01,fra10}; (iii) studying the influence of the
finite sizes of the beam in the longitudinal and the transversal directions on the instability. An expected effect is the
self--consistent generation of the counterstreaming current in a plasma \cite{hon00} which is neglected in the present
study. However, this is not a principal restriction on our treatment and the return current can be included by adding in
Eqs.~\eqref{eq:30} and \eqref{eq:31} the similar terms but with the flow velocity $\mathbf{u}_{e}$ \cite{bre04,bre05,bre06}.
The latter is determined from the condition of the beam current neutralization, $n_{e}\mathbf{u}_{e} \simeq -n_{b}\mathbf{u}_{b}$;
(iv) another important issue not considered here is the effect of the RF on the dynamics of the beam particles. This implies
either relativistic beams (with $\gamma_{b}\gg 1$) or nonrelativistic beams of heavy particles (heavy ions, protons,
antiprotons etc.); (v) considering the contributions of the higher harmonics with $\ell\geqslant 2$. From the structures
of the equations~\eqref{eq:39}, \eqref{eq:40} and \eqref{eq:47}--\eqref{eq:48} it follows that in this case the combined
frequencies $\ell \omega_{0}\pm \omega_{p}$ come into play determining some new branches for the unstable modes. In principle,
the study of the generation of the higher harmonics of these modes could be facilitated employing a dynamical equation for
the complex amplitude similar to Eq.~\eqref{eq:bp5}; (vi) lastly, studying some other orientations of the polarization
vector $\mathbf{E}_{0}$ of the RF and the beam drift velocity $\mathbf{u}_{b}$ with respect to $\mathbf{k}$. However, in
general, the simultaneous investigation of these issues is a formidable task and requires several separate investigations.
We intend to address these issues in our forthcoming investigations.

\begin{acknowledgments}
One of the authors, H.B.N., is grateful for the financial support of the Universit\'{e} Paris--Sud XI, Orsay, and
thanks the staff of the Plasma Physics Laboratory for their hospitality during his visit. Also the work of H.B.N.
has been partially supported by the State Committee of Science of Armenian Ministry of Higher Education and Science
(Project No. 11--1c317).
\end{acknowledgments}

\appendix

\section{Beam--plasma modes in a cold--fluid approximation}
\label{sec:app0}

In this Appendix in a cold--fluid model we briefly consider the asymptotic
behavior of the frequencies and the growth rates of the standard (in the
absence of the RF, $v_{E}=0$) beam--plasma longitudinal modes at large and
small $k$. Similar qualitative analysis have been conducted in Ref.~\cite%
{mik74}. We look for the solutions of the dispersion equation $\mathcal{D}%
_{\parallel }(k,\omega )=0$ of the longitudinal modes in the form $\omega
=\omega _{r}+i\gamma $, where $\omega _{r}$\ is the (real) frequency and $%
\gamma $ is the growth rate of the modes, respectively. Here the dispersion
function $\mathcal{D}_{\parallel }(k,\omega )$\ is determined by Eq.~%
\eqref{eq:32} with Eqs.~\eqref{eq:31} and \eqref{eq:cold}.

As has been mentioned in Sec.~\ref{sec:4.1} at short wavelengths, $k\geqslant k_{c}$,
there are four real solutions (i.e. $\gamma (k)=0$) of the dispersion equation which
at $k\geqslant k_{c}$ asymptotically behave as
\begin{eqnarray}
&&\omega _{r1}^{\pm }( k) =\pm \omega _{p}\left[ 1+\frac{\xi
\omega _{p}^{2}}{2\gamma _{b}^{3}( ku_{b}) ^{2}}\pm \frac{\xi
\omega _{p}^{3}}{\gamma _{b}^{3}( ku_{b}) ^{3}}+...\right] ,
\label{eq:cp1} \\
&&\omega _{r2}( k) =ku_{b}\left[ 1-\frac{\sqrt{\xi }\omega _{p}}{%
\gamma _{b}^{3/2}ku_{b}}-\frac{\sqrt{\xi }\omega _{p}^{3}}{2\gamma
_{b}^{3/2}( ku_{b}) ^{3}}+...\right] ,  \label{eq:cp2} \\
&&\omega _{r3}( k) =ku_{b}\left[ 1+\frac{\sqrt{\xi }\omega _{p}}{%
\gamma _{b}^{3/2}ku_{b}}+\frac{\sqrt{\xi }\omega _{p}^{3}}{2\gamma
_{b}^{3/2}( ku_{b}) ^{3}}+...\right] .  \label{eq:cp3}
\end{eqnarray}

The modes with the frequencies $\omega _{r3}(k)$ and $\omega _{r1}^{-}(k)$
remain stable also at $0\leqslant k<k_{c}$ while the other mode with $\omega
_{r1}^{+}(k)$ at $k=k_{c}$ merges with the mode $\omega _{r2}(k)$. The
latter becomes unstable at $k<k_{c}$. The frequency and the growth rate of
this mode in this long wavelength limit ($k<k_{c}$) asymptotically behave as
\begin{eqnarray}
&&\omega _{r2}( k) =\frac{ku_{b}}{1+\xi /\gamma _{b}^{3}}\left[ 1+%
\frac{4\xi ( 1-\xi /\gamma _{b}^{3}) }{\gamma _{b}^{3}(
1+\xi /\gamma _{b}^{3}) ^{3}}\frac{( ku_{b}) ^{2}}{\omega
_{p}^{2}}+...\right] ,  \label{eq:cp4} \\
&&\gamma ( k) =ku_{b}\frac{( \xi /\gamma _{b}^{3})
^{1/2}}{1+\xi /\gamma _{b}^{3}}\left[ 1+\frac{1-6\xi /\gamma _{b}^{3}+\xi
^{2}/\gamma _{b}^{6}}{( 1+\xi /\gamma _{b}^{3}) ^{3}}\frac{(
ku_{b}) ^{2}}{\omega _{p}^{2}}+...\right] .  \label{eq:cp5}
\end{eqnarray}

Similarly one finds the asymptotic behavior of the stable modes $\omega
_{r1}^{-}(k)$\ and $\omega _{r3}(k)$ at $k<k_{c}$,
\begin{equation}
\left(
\begin{array}{c}
\omega _{r1}^{-}( k)  \\
\omega _{r3}( k)
\end{array}%
\right) =\mp \omega _{p}\sqrt{1+\xi /\gamma _{b}^{3}}\left[ 1\mp \frac{\xi
/\gamma _{b}^{3}}{( 1+\xi /\gamma _{b}^{3}) ^{3/2}}\frac{ku_{b}}{%
\omega _{p}}+\frac{3\xi /\gamma _{b}^{3}}{2( 1+\xi /\gamma
_{b}^{3}) ^{3}}\frac{( ku_{b}) ^{2}}{\omega _{p}^{2}}+...%
\right] .  \label{eq:cp6}
\end{equation}

It is seen that the leading terms in Eqs.~\eqref{eq:cp4} and \eqref{eq:cp5}
coincide with $\omega _{\parallel }^{\mathrm{TS}}(k)$ and $\gamma ^{\mathrm{%
TS}}(k)$, respectively, see Eq.~\eqref{eq:TS}.

\section{Evaluation of the sum}
\label{sec:app1}

In this Appendix within fluid approximation we briefly derive an analytic expression for the function
$\Psi _{\ell }(\zeta )$ introduced in Sec.~\ref{sec:5} (see Eq.~\eqref{eq:psi}). Inserting Eq.~\eqref{eq:cold}
into Eq.~\eqref{eq:psi} we arrive at
\begin{equation}
\Psi _{\ell }( \zeta ) =( -1) ^{\ell }\frac{\omega _{p}%
}{2\omega _{0}}\sum_{s=-\infty }^{\infty }\left( \frac{1}{s+a_{+}}-\frac{1}{%
s+a_{-}}\right) J_{s}( \zeta ) J_{s+\ell }( \zeta ) ,
\label{eq:ap1}
\end{equation}%
where $a_{\pm }=(\omega \pm \omega _{p})/\omega _{0}$. The summation in Eq.~\eqref{eq:ap1} can be easily done
using the Newberger's summation formula \cite{new82} which is valid for noninteger $\mu $, $0\leqslant \gamma
\leqslant 1$ and $\mathrm{Re}(\alpha +\beta )>-1$,
\begin{equation}
\sum_{n=-\infty }^{\infty }( -1) ^{n}\frac{J_{\alpha -\gamma
n}( \zeta ) J_{\beta +\gamma n}( \zeta ) }{n+\mu }=%
\frac{\pi }{\sin ( \pi \mu ) }J_{\alpha +\gamma \mu }( \zeta
) J_{\beta -\gamma \mu }( \zeta ) .
\label{eq:ap2}
\end{equation}%
In general $\mu $, $\alpha $ and $\beta $ are complex quantities. Using the
summation formula \eqref{eq:ap2} from Eq.~\eqref{eq:ap1} at $\ell \geqslant
0 $ and $\ell \leqslant 0$ we obtain
\begin{equation}
\Psi _{\ell }( \zeta ) =( -1) ^{\ell }\frac{\pi \omega
_{p}}{2\omega _{0}}\left[ \frac{1}{\sin ( \pi a_{+}) }%
J_{a_{+}}( \zeta ) J_{\ell -a_{+}}( \zeta ) -\frac{1}{%
\sin ( \pi a_{-}) }J_{a_{-}}( \zeta ) J_{\ell
-a_{-}}( \zeta ) \right] ,
\label{eq:ap3}
\end{equation}%
\begin{equation}
\Psi _{\ell }( \zeta ) =\frac{\pi \omega _{p}}{2\omega _{0}}\left[
\frac{1}{\sin ( \pi a_{+}) }J_{a_{+}-\ell }( \zeta )
J_{-a_{+}}( \zeta ) -\frac{1}{\sin ( \pi a_{-}) }%
J_{a_{-}-\ell }( \zeta ) J_{-a_{-}}( \zeta ) \right] ,
\label{eq:ap4}
\end{equation}%
respectively.

Consider also the limit of the function $\Psi _{\ell }(\zeta )$ at small parameter $\zeta $ which correspond to the
limit of a weak RF. At $\zeta \ll 1$ using the asymptotic behavior of the Bessel function at small argument \cite{gra80}
from Eqs.~\eqref{eq:ap3} and \eqref{eq:ap4} we obtain
\begin{equation}
\Psi _{\ell }(\zeta ) =(-1) ^{\frac{\ell +\vert \ell \vert }{2}}\left( \frac{\zeta }{2}\right) ^{\vert \ell
\vert }\sum_{s=0}^{\vert \ell \vert }\frac{(-1)^{s}}{s!(\vert \ell \vert -s)!}\frac{\delta
\varepsilon _{\parallel }( -\eta _{\ell }s) }{1+\delta\varepsilon _{\parallel }(-\eta _{\ell }s)}+\mathrm{O}\left(
\zeta ^{\left\vert \ell \right\vert +2}\right) ,
\label{eq:ap5}
\end{equation}%
where $\eta _{\ell }=\vert\ell \vert/\ell $.

\section{Dynamical equation for the complex amplitudes}
\label{sec:app2}

In Sec.~\ref{sec:5}, we have derived the dispersion equation for the plasma
modes generated simultaneously by the laser radiation and the REB. For some applications the derivation of a differential
equation for the complex amplitude of the excited wave is desirable. We
consider here the case of a cold plasma when the partial dielectric function
is given by Eq.~\eqref{eq:cold}. For deriving the dynamical equation for the
amplitude, we insert Eqs.~\eqref{eq:cold} and \eqref{eq:psi} into Eq.~%
\eqref{eq:47} and using the summation formula \eqref{eq:sum} we represent
the latter in the form
\begin{equation}
E_{\parallel }(0) =-\sum_{\ell =-\infty }^{\infty }E_{\parallel
}(\ell ) \delta \varepsilon _{b\parallel }(-\ell )
\sum_{s=-\infty }^{\infty }\frac{(\omega -s\omega _{0}) ^{2}}{%
(\omega -s\omega _{0})^{2}-\omega _{p}^{2}}J_{s-\ell }(\zeta ) J_{s}(\zeta ) .
\label{eq:st}
\end{equation}%
Now, in both sides of Eq.~\eqref{eq:st} we make an inverse Fourier
transformation. Then denoting the time--dependent complex amplitude as $%
\mathcal{E}(t)$ and using the summation formula \cite{gra80}
\begin{equation}
\sum_{\ell =-\infty }^{\infty }e^{\pm i\ell \omega _{0}t}J_{\ell }(\zeta ) =e^{\pm i\zeta \sin (\omega _{0}t)},
\label{eq:bp1}
\end{equation}%
we obtain an equation
\begin{equation}
\mathcal{E}(t) e^{i\zeta \sin (\omega _{0}t)
}=-\int_{-\infty }^{\infty }d\omega E_{\parallel }\left( \omega \right)
\delta \varepsilon _{b\parallel }\left( \omega \right) \sum_{s=-\infty
}^{\infty }\frac{\left( \omega -s\omega _{0}\right) ^{2}}{\left( \omega
-s\omega _{0}\right) ^{2}-\omega _{p}^{2}}e^{-i\left( \omega -s\omega
_{0}\right) t}J_{s}\left( \zeta \right) .
\label{eq:bp2}
\end{equation}%
Next, on both sides of Eq.~\eqref{eq:bp2} we apply the differential operator
$\partial ^{2}/\partial t^{2}+\omega _{p}^{2}$,
\begin{eqnarray}
&&\left( \frac{\partial ^{2}}{\partial t^{2}}+\omega _{p}^{2}\right) \left[
\mathcal{E}\left( t\right) e^{i\zeta \sin \left( \omega _{0}t\right) }\right]
=-\frac{\partial ^{2}}{\partial t^{2}}e^{i\zeta \sin \left( \omega
_{0}t\right) }\int_{-\infty }^{\infty }\delta \varepsilon _{b\parallel
}\left( \omega \right) E_{\parallel }\left( \omega \right) e^{-i\omega t}d\omega   \label{eq:bp3} \\
&=&\frac{\omega _{b}^{2}}{\gamma _{b}^{3}}\frac{\partial ^{2}}{\partial t^{2}%
}e^{i\zeta \sin \left( \omega _{0}t\right) }\int_{-\infty }^{\infty }%
\mathcal{E}\left( \tau \right) G_{b}\left( \tau -t\right) d\tau .  \nonumber
\end{eqnarray}%
In the last part of Eq.~\eqref{eq:bp3} we have used the longitudinal dielectric function of the beam, Eq.~\eqref{eq:31}.
Here $G_{b}(t)$ is the Green function of the electron beam
\begin{equation}
G_{b}\left( t\right) =\frac{1}{2\pi }\int_{-\infty }^{\infty }\frac{%
e^{i\omega t}d\omega }{(\omega -ku_{b} +i0)^{2}} ,
\label{eq:bp4}
\end{equation}%
where we have introduced the positive infinitesimal $+i0$ which guarantees the causality of the response. An explicit
expression for the function $G_{b}(t)$ can be easily obtained employing in Eq.~\eqref{eq:bp4} the contour integration
technique. The result reads as $G_{b}(t)=\Theta (-t)te^{iku_{b}t}$, where $\Theta (t)$ is the Heaviside unit--step
function. Substituting this result into Eq.~\eqref{eq:bp3} we finally arrive at
\begin{equation}
\left( \frac{\partial ^{2}}{\partial t^{2}}+\omega _{p}^{2}\right) \left[
\mathcal{E}\left( t\right) e^{i\zeta \sin \left( \omega _{0}t\right) }\right]
=\frac{\omega _{b}^{2}}{\gamma _{b}^{3}}\frac{\partial ^{2}}{\partial t^{2}}%
e^{i\zeta \sin \left( \omega _{0}t\right) }\int_{-\infty }^{t}\mathcal{E}%
\left( \tau \right) e^{iku_{b}\left( \tau -t\right) }\left( \tau
-t\right) d\tau .
\label{eq:bp5}
\end{equation}%
The obtained equation \eqref{eq:bp5} must be accompanied by the initial conditions. When these conditions are
specified Eq.~\eqref{eq:bp5} represents the evolution of the complex amplitude of the waves excited simultaneously
by the RF and the REB in a cold plasma. Note that in contrast to the dispersion equation
\eqref{eq:49}, Eq.~\eqref{eq:bp5} is also valid for the intensity parameter $\zeta\simeq ka\sim 1$.

\end{document}